# ESG Beliefs of Large Language Models: Evidence and Impact

Tong Li[*]          Luping Yu[†]

December 2025

**Abstract**

We examine whether large language models (LLMs) hold systematic beliefs about environmental, social, and governance (ESG) issues and how these beliefs compare with—and potentially influence—those of human market participants. Based on established surveys originally administered to professional and retail investors, we show that major LLMs exhibit a strong pro-ESG orientation. Compared with human investors, LLMs assign greater financial relevance for ESG performance, expect larger return premia for high-ESG firms, and display a stronger willingness to sacrifice financial returns for ESG improvements. These preferences are highly uniform and values-driven, in contrast to heterogeneous human views. Using a large dataset of analyst reports, we further show that sell-side analysts become significantly more optimistic about high-ESG firms after adopting LLMs for research. Our findings reveal that LLMs embed distinct, coherent ESG beliefs and that these beliefs can shape human judgments, highlighting a new channel through which AI adoption may influence financial markets.

*JEL Classifications*: G11, G14, G24, Q56
*Keywords*: Large Language Models, ESG, Sustainable Finance, Surveys, Financial Analysts

[*] School of Economics & Wang Yanan Institute for Studies in Economics, Xiamen University. Email: litong@xmu.edu.cn.
[†] Faculty of Business Administration, University of Macau. Email: lupingyu@um.edu.mo.

# 1 Introduction

The financial sector is undergoing a profound transformation as large language models (LLMs) become deeply integrated into investment research, corporate disclosure, and investor trading decisions. A growing body of evidence shows that investors now rely on LLMs to interpret news and generate trade ideas (Cheng, Lin, and Zhao, 2025; Chang, Dong, Martin, and Zhou, 2025; Sheng, Sun, Yang, and Zhang, 2025); firms increasingly turn to these models to draft public filings (Blankespoor, deHaan, and Li, 2025); and sell-side equity analysts start using them to process information and generating research outputs (Bertomeu, Lin, Liu, and Ni, 2025; Christ, Kim, and Yip, 2025). Research further demonstrates that AI systems can digest vast volumes of qualitative and quantitative information and generate stock-return forecasts that outperform most human experts, while offering insights that meaningfully complement human judgment (Cao, Jiang, Wang, and Yang, 2024). These developments suggest that LLMs are no longer peripheral tools but active decision-shaping inputs in financial markets. As AI capabilities expand and AI-assisted decisions become increasingly prevalent, a set of foundational questions emerges: Do LLMs exhibit systematic beliefs and preferences toward value-relevant topics? If so, do these beliefs resemble those held by human market participants? And to what extent do these model-embedded beliefs shape the judgments and behaviors of market participants?

We address these important questions by focusing on beliefs and preferences related to environmental, social, and governance (ESG) issues, a domain that has become a mainstream pillar of global investing and sparked intense academic and policy debate. Over the past decade, ESG considerations have become deeply incorporated into asset-management practices, shareholder engagement, regulatory frameworks, and corporate decision-making. Yet investors continue to exhibit wide dispersion, and in some cases confusion, about the financial relevance, ethical meaning, and portfolio implications of ESG factors (Starks, 2023). This combination of rising importance and persistent disagreement makes ESG an ideal setting to study how LLMs internalize complex, value-relevant concepts and whether their beliefs align with or diverge from those of human decision-makers. Moreover, because sustainability preferences can influence firm valuations, capital allocation, and corporate behavior, any systematic bias or tilt in LLM-embedded ESG beliefs may have far-reaching consequences as these models increasingly shape decisions of market participants. Examining ESG beliefs therefore provides a clear and consequential lens for understanding not only what LLMs "think," but also how those beliefs may influence financial markets via the humans who rely on them.

To elicit the ESG beliefs embedded in large language models, we adapt two benchmark survey frameworks originally designed for human investors—Edmans et al. (2025) for professional equity portfolio managers and Giglio et al. (2025) for U.S. retail investors. These instruments provide structured, quantitative measures of how individuals evaluate ESG materiality, return expectations, investment motives, and climate-related attitudes. To ensure that AI-generated responses are directly comparable to human survey data, we employ a role-definition prompting strategy. Each model is instructed to adopt a realistic professional identity and informational frame before answering. For instance, when responding to Edmans et al. (2025), the model assumes the role of an active equity portfolio manager operating between November 2023 and February 2024. Prior work shows that such contextual identity priming aligns AI responses more closely with human reasoning, improving the internal validity of cross-group comparisons (e.g., Fedyk, Kakhbod, Li, and Malmendier, 2025).

We implement this protocol across a panel of ten frontier LLMs representing major global providers, including OpenAI, Anthropic, Google, DeepSeek, and Mistral, with multiple generations of each model family. For every model-question pair, we generate 100 independent responses under low-randomness configurations, which induce deterministic behavior conditional on the prompt. Averaging these 100 draws yields each model's representative stance while filtering out stochastic noise. This repeated-elicitation design, coupled with cross-model and cross-generation variation, produces a stable and comparable measure of underlying ESG orientations. It allows us to identify systematic patterns in how AI systems evaluate ESG relevance and long-term financial performance. Taken together, this methodology provides a robust empirical foundation for comparing LLM-embedded ESG beliefs with those documented among human investors in prior large-scale surveys, thereby linking algorithmic reasoning to established evidence on human sustainability preferences.

We begin our experimental analysis with the survey of global active equity portfolio managers conducted by Edmans et al. (2025). The first question asks managers to rank the importance of actual environmental and social (ES) performance for long-term firm value relative to five other value drivers: strategy and competitive position, operational performance, corporate culture, governance, and capital structure. The authors document that professional investors, including self-identified sustainable investors, on average place ES performance at the bottom of these six determinants. By contrast, we find that every major LLM assigns substantially greater importance to ES considerations, positioning them much closer to the core of firm valuation. On average across AI models, ES performance ranks fifth, significantly above capital structure and comparable to corporate culture. This elevated ranking is highly stable across developers, even those operating in different countries, suggesting that geographic origin does not meaningfully shape how LLMs assess ESG materiality. The consistency in LLM responses is in sharp contrast to the wide dispersion of beliefs observed among human portfolio managers. Taken together, these findings show that LLMs encode a distinct and homogenized "ESG tilt": they systematically attribute greater importance to sustainability considerations than human investors and do so with far less heterogeneity.

We continue to examine how LLMs assess the financial materiality of specific environmental and social dimensions. Across all eight ES categories in Edmans et al. (2025), LLMs assign substantially higher materiality than human portfolio managers. Whereas the average materiality rating among fund managers is 2.29 on a 0–4 scale, the AI-wide average rises to 2.84, indicating a pronounced upward shift in perceived financial relevance. The magnitude of this AI–human gap varies meaningfully across categories. Environmental topics, particularly greenhouse gas emissions and ecological impacts, exhibit the largest increases, with AI models rating these dimensions far more financially consequential than human respondents (e.g., 3.41 vs. 2.50 for emissions; 3.16 vs. 2.23 for ecological impacts). Upward adjustments are more modest for social issues such as demographic diversity or supplier treatment, where human and AI assessments are more closely aligned.

Large language models also differ substantially from human portfolio managers in their expectations about the future stock-market performance of firms with strong versus weak environmental and social (ES) records. In Edmans et al. (2025), professional investors anticipate only modest performance differentials: they expect good ES firms to slightly outperform (mean = 0.57 on a –2 to +2 scale) and poor ES firms to slightly underperform (–0.70). By contrast, LLMs embed considerably more polarized and internally consistent beliefs. On average, AI models predict a much wider spread of +1.19 for good ES firms and –1.22 for poor ES firms, implying both a stronger ESG-return premium and a more pessimistic assessment of ESG laggards. This pattern is remarkably consistent across all ten models: good ES performers are uniformly expected to outperform, while poor ES performers are expected to underperform. Most models, including GPT-5, Claude-4, Gemini 2.0, and DeepSeek, cluster tightly around ±1, indicating moderate but persistent long-term return differentials, whereas a few (e.g., GPT-4 and Gemini 2.5) generate more extreme "strongly outperform" or "strongly underperform" predictions. This broad alignment among models contrasts sharply with the heterogeneity among human respondents: in Edmans et al. (2025), only 9% of portfolio managers believe good ES firms will "strongly outperform," and responses span the full –2 to +2 range. Overall, LLMs not only attribute greater materiality to ESG factors but also exhibit stronger and more uniform expectations about their financial payoffs, reinforcing the systematic "ESG tilt" of AI models.

We also examine how LLMs balance financial performance against ESG considerations when explicit trade-offs arise. Theoretically, ESG preferences have well-defined financial implications: investors with stronger sustainability tastes may rationally accept lower expected returns in exchange for holding greener assets (Pástor et al., 2021). In practice, however, human portfolio managers in Edmans et al. (2025) generally resist sacrificing returns for sustainability. One-third report unwillingness to accept any long-term performance reduction, a view shared even by more than 20% of sustainable fund managers. Only 9% of the respondents are willing to tolerate a loss of 11–50 basis points per year to improve ES outcomes. By contrast, LLMs consistently display greater willingness to accept financial sacrifices in exchange for ESG improvements. Across ten models, none select the "no-sacrifice" option; instead, they indicate willingness to accept at least a modest trade-off, with 47% (17%) in the 1–10 (11–50) basis-point range. Several models, such as Claude-3 and Gemini 2.0, even respond that meaningful ESG gains justify return reductions of 11–50 basis points annually. The modal AI judgment thus departs from the human baseline, implying that genuine trade-offs between sustainability and financial goals are both necessary and acceptable. The evidence indicates that LLMs exhibit stronger nonpecuniary preferences for ESG investing—a values-driven stance in the spirit of Starks (2023).

We complement these findings by comparing LLM-embedded ESG beliefs with those of U.S. retail investors surveyed in Giglio et al. (2025), and again observe a pronounced and systematic divergence. Human investors exhibit substantial heterogeneity in their expectations, motives, and climate-related attitudes; LLMs, by contrast, converge toward a far more uniform and consistently pro-ESG set of beliefs. For return expectations, retail investors anticipate a sizable performance gap in favor of the broad U.S. stock market: 7.13% versus 5.20% annually for diversified ESG portfolios over a 10-year horizon. The AI-wide average market expectation is slightly more conservative than the human benchmark (6.80%), yet the average expected return for ESG portfolios increases to 6.60%, closing the human-implied underperformance gap of nearly 200 basis points to just 20 basis points in annualized terms. Moreover, LLMs display dramatically lower dispersion in expected returns: standard deviations of 0.7–0.9 percentage points, compared with 4–5 percentage points for human investors. Overall, LLMs exhibit a highly homogenized "consensus view" in which ESG portfolios are perceived as delivering long-term performance almost on par with the overall market.

LLMs also express markedly different motivations for ESG investing. While 48% of retail investors in Giglio et al. (2025) report "no specific reason" for holding ESG assets, LLMs select "it is the right thing to do" as the dominant explanation. On average, 61% of AI responses cite this ethical rationale, more than doubling the human share, while 39% identify climate-risk hedging. Not a single model chooses "no specific reason" or "ESG will outperform." This sharp contrast again indicates that LLMs interpret ESG investing primarily through a moral or values-driven lens, rather than as a financially motivated decision. A consistent pattern emerges with climate-change concerns. Human beliefs are balanced across low, moderate, and high concern; by contrast, every LLM assigns itself to the "high concern" category. Across AI models, roughly 18% classify themselves as "extremely concerned" and the remaining 82% as "very concerned." This consistency reflects a strong, internalized conviction that climate risk is a critical, first-order investment consideration. Together, these retail-investor comparisons reinforce the broader patterns observed throughout our analysis. Relative to the heterogeneous and often weakly articulated beliefs of human investors, LLMs encode more uniform, assertive, and deeply values-driven ESG beliefs that consistently place sustainability, climate risk, and ethical motives at the forefront of long-horizon investment decision-making.

A natural question raised by these systematic differences in ESG beliefs is whether the "ESG tilt" embedded in large language models manifests in real financial decision-making. This study focuses on sell-side equity analysts, whose assessments guide the decisions of a wide range of capital market participants (e.g., Womack, 1996; Howe, Unlu, and Yan, 2009; Lee, Ng, and Swaminathan, 2009), and who increasingly rely on LLMs to draft, edit, and refine their research reports (Bertomeu, Lin, Liu, and Ni, 2025). Accordingly, we investigate whether and how AI-embedded ESG beliefs influence the judgement of financial analysts. Using a comprehensive dataset of more than 36,000 analyst reports from top brokerage firms in 2023 and 2024, linked to analysts' historical stock recommendation records and firms' ESG performance metrics, we identify the presence and extent of AI-generated text through a state-of-the-art AI detection model. This approach allows reports to be classified by the degree of AI involvement and enables observation of changes in analysts' behavior before and after adopting LLM tools.

We then examine whether analysts who incorporate LLMs into their workflow implicitly import the models' pro-ESG priors into their stock analyses. Multiple pieces of evidence appear to support this conjecture. First, analyst reports with AI-generated content devote substantially more attention to ESG-related topics. Second, after adopting AI tools, analysts are more likely to issue positive recommendations, including buy or strong buy ratings and recommendation upgrades, for firms with higher ESG scores. Third, analysts who use large language models become significantly more optimistic about high-ESG firms relative to the overall analyst consensus, but these AI-adopting analysts exhibit slightly lower optimism toward low-ESG firms. In other words, LLMs do not make analysts consistently more bullish; instead, their influence is concentrated on firms whose sustainability performance aligns with the models' internal ESG beliefs. These findings suggest that the ESG preferences embedded in large language models can meaningfully shape the judgments of financial intermediaries.

Our study contributes to the literature on ESG beliefs and perceptions by introducing a fundamentally new class of decision-making agents—large language models—into a body of work that has, until now, focused exclusively on human actors. Prior research documents substantial heterogeneity in ESG perceptions across institutional investors, retail investors, and financial analysts, with differences in motives, return expectations, and willingness to sacrifice performance for sustainability (e.g., Krueger, Sautner, and Starks, 2020; Chen, Li, Ma, and Michaely, 2025; Edmans et al., 2025; Giglio et al., 2025). Yet all existing evidence centers on human beliefs, leaving open the question of whether AI systems, now increasingly employed in financial analysis, internalize systematic ESG beliefs of their own. We fill this gap by systematically eliciting ESG beliefs from frontier LLMs under both professional and retail investor identities based on established survey instruments. Our analysis provides novel evidence that LLMs exhibit stable and internally consistent ESG preferences that are substantially more pro-ESG than those of human investors.

We further contribute to the emerging literature comparing preferences revealed by humans and AI-generated agents. Recent research shows that large language models can exhibit stable and internally coherent preference structures, which are sometimes more consistent or more extreme than those observed in humans. For example, studies of economic rationality (e.g., Chen, Liu, Shan, and Zhong, 2023) demonstrate that LLMs make decisions that largely conform to utility maximization and can adapt recommendations to reflect human risk preferences. Other work compares AI and human judgments in applied domains such as investment decisions, ethics, and auditing (e.g., Bertomeu, Cheynel, Lunawat, and Milone, 2025; Fedyk et al., 2025; MacKenzie et al., 2025), revealing that while LLMs often mirror human reasoning, they can also encode systematic biases or exhibit distinct patterns of judgment. A complementary strand of research treats LLMs as simulable agents for in-silico social-science experiments (e.g., Manning, Zhu, and Horton, 2024; Zarifhonarvar and Ganji, 2025), further documenting coherent and context-responsive preference patterns embedded within these models. Notably, this literature has focused on domains such as risk-taking, ethical reasoning, political attitudes, and personnel decisions. No prior study has examined whether LLMs internalize coherent beliefs about ESG, how those beliefs compare with those of human investors, or whether they shape the judgments of AI users. By addressing these questions, our paper extends the LLM-preferences literature into a domain with direct and growing relevance for asset pricing, corporate behavior, and sustainable development.

Our study also extends the rapidly growing literature on how AI adoption reshapes financial markets. Prior work shows that AI-based analysts can outperform most human analysts when information is transparent but voluminous, while human experts retain advantages in settings requiring institutional knowledge, and human-AI collaboration meaningfully reduces extreme forecasting errors (Cao et al., 2024). In addition, the access to ChatGPT directly improves the work of financial intermediaries and enhances information environments in capital markets (Bertomeu, Lin, Liu, and Ni, 2025). A few studies further suggest that investors themselves rely on LLMs for trading. Specifically, ChatGPT outages depress trading volume, price impact, and intraday volatility (Cheng et al., 2025). Chang et al. (2024) suggest that the diffusion of ChatGPT enables retail traders to align more closely with AI-derived signals, reducing informational disparities with short sellers. On the other hand, Sheng et al. (2025) document the widespread use of ChatGPT among hedge funds and show that generative AI, which brings superior returns to adopters, could widen existing disparities among investors. Together, this literature establishes that AI use already shapes trading intensity, intermediary behavior, and price formation. Yet it focuses almost exclusively on how AI aids decision-making, rather than examining the beliefs embedded within AI systems or how those beliefs may feed back into human judgment. By eliciting ESG-related beliefs directly from LLMs and tracing their influence on sell-side analysts' recommendations, our study fills this gap and introduces a new perspective for understanding the financial-market consequences of LLM adoption.

## 2 Background and Related Literature

### 2.1 ESG Beliefs and Perceptions in Financial Markets

A large and growing literature examines how market participants perceive and incorporate ESG considerations into financial decision-making. For example, Krueger et al. (2020) survey various types of participants, including institutional investors, senior executives, analysts, and ESG specialists, and they show that many integrate climate considerations due to reputational, ethical, or stakeholder pressures rather than purely financial motivations. Edmans et al. (2025)

broaden the lens by studying a wider set of ES issues and explicitly targeting the central question of sustainable investing: whether improving societal outcomes aligns with or conflicts with shareholder value. They find that, unlike climate risk, which is often incorporated for non-financial reasons, ES performance more generally is viewed predominantly through a financial-materiality lens.

Several studies focus on the ESG perceptions of retail investors. Giglio et al. (2025) show that U.S. retail investors generally expect ESG stocks to underperform, and ESG-oriented investors tend to be motivated by ethical considerations or climate-hedging motives rather than return maximization. These findings echo evidence from Riedl and Smeets (2017), who match survey responses to portfolio holdings of Dutch investors and show that non-pecuniary motives play a central role in sustainable investment choices. Other studies further examine ESG perceptions among private equity and venture capital investors (McCahery, Pudschedl, and Steindl, 2022), finance professionals involved in valuation (Bancel, Glavas, and Karolyi, 2025), and market participants' expectations about climate risk and return (Bauer, Gödker, Smeets, and Zimmermann, 2024). Overall, this literature reveals substantial heterogeneity in how different types of participants understand and prioritize ESG considerations.

Another related strand of the literature examines whether investors are willing to pay or sacrifice returns for sustainability. For example, in a panel of experienced private investors and dedicated high-net-worth impact investors, Heeb, Kölbel, Paetzold, and Zeisberger (2023) document a substantial willingness-to-pay for sustainable impact but the magnitude is not sensitive to the level of impact. This finding is also confirmed in a larger sample of individual participants. Baker Egan, and Sarkar (2024) estimate that investors are willing to pay roughly 20 basis points in fees for ESG fund mandates. These findings complement a broader set of survey and experimental studies documenting a positive willingness to pay for sustainability or impact (e.g., Humphrey, Kogan, Sagi, and Starks, 2021; Bauer et al., 2024; Engler, Gutsche, and Smeets, 2024), as well as evidence from fund flows (e.g., Döttling and Kim, 2024).

Finally, recent work examines ESG perceptions among financial analysts. Chen et al. (2025) show that sell-side analysts prioritize shareholder value maximization even in environmental contexts, reinforcing the view that market participants primarily evaluate ESG through a financial lens. Because analyst perspectives are often regarded as a proxy for market opinions more broadly (e.g., Brav, Lehavy, and Michaely, 2005; Pástor, Sinha, and Swaminathan, 2008; Bordalo, Gennaioli, La Porta, and Shleifer, 2019), their findings highlight the centrality of financial value in expert ESG assessments. Chen et al. (2025) also emphasize the role of "environmental opportunities," expanding the traditional focus on environmental risks and offering a more nuanced perspective on how experts evaluate ESG-related value creation.

Across these studies, a clear message is that ESG beliefs are heterogeneous, contested, and consequential for financial behavior. However, existing studies only focus on human perceptions. No prior work investigates whether large language models, now gradually integrated into financial analysis and decision-making, internalize systematic ESG beliefs or how these beliefs compare to those of human investors. Likewise, no study examines whether AI-embedded preferences influence the judgments of market participants who rely on these tools. Our study fills this critical gap by (i) systematically eliciting ESG beliefs embedded in frontier LLMs using validated human survey instruments, and (ii) documenting how these model-encoded beliefs propagate into the decisions of financial analysts. In doing so, we extend the literature on ESG perceptions to a new and increasingly influential actor in modern financial markets: artificial intelligence.

## 2.2 Perceptions and Biases of Large Language Models

State-of-the-art large language models, such as OpenAI's GPT, Anthropic's Claude, and Google's Gemini, are trained on vast and diverse internet corpora that encompass a wide range of viewpoints. However, these datasets are often disproportionately dominated by liberal or progressive narratives. For instance, OpenAI's GPT-3 training set reportedly includes about 60% web-scraped content, which is inherently susceptible to biases stemming from the dominance of certain ideological perspectives on the internet. As a result, the underlying content these models are trained on may reflect specific political, social, and cultural viewpoints, often aligned with progressive ideals.

In addition to these base datasets, LLMs undergo a final stage of training known as Reinforcement Learning from Human Feedback (RLHF), which is designed to further refine model outputs by aligning them with "human preferences" and ethical guidelines. This stage is particularly influential in shaping how models respond to morally and politically charged topics. However, RLHF also introduces potential biases, as the feedback and ethical guidelines provided by human annotators, many of whom are drawn from the tech sector, can carry their own value orientations. As OpenAI CEO Sam Altman has acknowledged, one of the risks of this process is the bias of human labelers, and that the homogeneity of their views can lead to the hardwiring of group biases into the model. Consequently, LLMs may inherit, or even amplify, the values held by their creators and the broader tech community, which often align with progressive views on issues such as equality, environmental protection, and social justice.

Recent research investigates whether large language models exhibit systematic preferences, biases, and decision patterns, as well as how these compare to human behavior. Several studies assess the economic rationality of LLMs. Chen et al. (2023) show that ChatGPT models make choices across risk, time, social, and food domains that are largely consistent with utility maximization and, in many cases, more internally coherent than those of human subjects. Kim, Kovach, Lee, and Shin (2024) replicate classic experiments on risk preferences and find that ChatGPT can learn and adapt its portfolio recommendations to match individuals' risk aversion, though it aligns less well with disappointment aversion. These papers collectively demonstrate that LLMs can express stable, structured preferences, while also revealing important divergences from human decision-making.

Other work examines whether LLM-generated judgments replicate or depart from human reasoning in applied domains. Fedyk et al. (2025) compare investment recommendations produced by AI-generated agents to those of over 1,200 human survey respondents. They find that default LLM responses disproportionately reflect the preferences of young, high-income individuals, indicating pronounced algorithmic bias. Yet such bias largely disappears once models are seeded with demographic context. Moreover, LLMs articulate investment rationales that closely mirror human explanations, and their responses can even help diagnose reasoning errors in human subjects. In domains involving nuanced ethical reasoning, Bertomeu, Cheynel, Lunawat, and Milone (2025) compare human and LLM judgments in a morally ambiguous reporting scenario. They show that LLMs apply ethical principles more consistently than humans but differ in how they interpret and justify moral trade-offs, raising questions about model alignment in value-sensitive tasks.

A complementary strand of work views LLMs as simulable agents capable of participating in large-scale, in silico social-science experiments. Manning et al. (2024) show that, when embedded within structural causal models, LLM-based agents can generate and test social-scientific hypotheses across settings such as negotiations, bail hearings, job interviews, and auctions. Crucially, these studies reveal that LLMs "know more than they can tell": elicited predictions about outcomes are often inaccurate, but once conditioned on a fitted structural causal model, the same LLMs generate responses that closely track theoretical predictions. This finding suggests that LLMs encode latent structural knowledge that becomes visible only when models are placed in environments that mimic causal decision-making processes, a result that deepens the interpretation of LLM behavior beyond surface-level prompt responses. This broader view of LLMs as implicit computational models of humans is also central to Horton (2023). They argue that LLMs can be deployed like homo silicus to simulate economic behavior and generate new hypotheses, as these models are trained on vast corpora of human-generated text. Similarly, Zarifhonarvar and Ganji (2025) demonstrate that LLMs exhibit coherent and context-responsive risk and time preferences, adjusting more systematically to macroeconomic shocks than smaller models and revealing preference patterns that diverge meaningfully from human subjects.

A growing body of work documents social and political biases embedded in LLM outputs. Westwood, Grimmer, and Hall (2025) use large-scale user evaluations to show that nearly all major LLMs are perceived as left-leaning across a broad set of political topics. Although prompting models to adopt a neutral stance reduces perceived partisanship, respondents across the political spectrum continue to identify systematic ideological slants. Fulay et al. (2024) document pronounced left-leaning tendencies in LLMs on issues like climate change and labor rights. An, Huang, Lin, and Tai (2025) uncover demographic biases in a high-stakes hiring context: GPT assigns higher scores to female candidates but systematically lower scores to Black male candidates with otherwise identical résumés. These patterns hold across job types and geographic subsamples, indicating that LLM decisions can replicate and sometimes amplify social biases reflected in training data. Research in auditing contexts reinforces this theme: MacKenzie et al. (2025) find that LLMs produce more conservative judgments than human auditors and do not always respond to experimental manipulations in the same way, suggesting differences in susceptibility to cognitive biases.

These studies suggest that LLMs systematically embed coherent preference structures, which may align with or be more extreme than human preferences, depending on the decision context. Existing research, however, has focused primarily on domains such as risk-taking, ethical reasoning, political expression, auditing, and personnel evaluation. Absent from this literature is any systematic investigation of whether LLMs internalize structured beliefs about ESG issues, or whether those embedded beliefs meaningfully shape the judgments of human market participants who increasingly rely on AI tools. Our study fills this gap by extending the analysis of LLM perceptions into an increasingly importantly domain in sustainable finance.

## 2.3 Financial Market Impact of Large Language Models

A rapidly expanding literature documents how the adoption of large language models and related AI tools reshape information production, trading behavior, and price formation in financial markets. Cao et al. (2024) highlights the transformative role of AI in financial analysis. Specifically, they develop an "AI analyst" trained on corporate disclosures, industry dynamics, and macroeconomic indicators. They show that the AI analyst outperforms most human analysts in predicting stock returns, particularly in environments where information is transparent but voluminous. Human analysts retain advantages in settings that rely on institutional knowledge, such as evaluating intangible assets or financial distress, where they are able to provide significant incremental value when combined with AI. Their findings demonstrate powerful human–machine complementarities: joint human-AI decision-making substantially reduces extreme forecasting errors, and analysts catch up to machines once firms build internal AI capabilities and gain access to alternative data.

In addition, Bertomeu, Lin, Liu, and Ni (2025) investigate how ChatGPT affects information intermediaries. Exploiting Italy's unexpected national ban on ChatGPT, they identify how analysts use AI in producing research outputs. Following the ban, domestic analysts issue fewer forecasts, rely more heavily on industry-level information, and produce less accurate predictions relative to foreign analysts covering the same firms. Market reactions to earnings announcements intensify and bid–ask spreads widen, pointing to a deterioration in the information environment. These effects are strongest for analysts whose pre-ban writing styles suggest heavier AI reliance and for those with technical training, highlighting heterogeneity in AI dependence.

Several studies examine how investors themselves use AI in trading. Cheng, Lin, and Zhao (2025) exploit exogenous ChatGPT outages and show that a substantial number of investors rely on generative AI for information processing: trading volume drops markedly during outages, especially for firms with recent news and those held by transient institutions. Consistent with reduced informed trading, they also document that both price impact and intraday return variance decline. AI-assisted trading also improves long-run price informativeness. Chang et al. (2024) highlight the democratizing effects of AI by constructing an AI-sentiment measure from earnings calls. Before ChatGPT's introduction, AI sentiment aligned with short sellers but not retail traders; afterward, alignment increases among retail traders and declines among short sellers. This pattern suggests that AI lowers information-processing frictions for retail investors, narrowing disparities between retail and sophisticated traders. On the other hand, Sheng et al. (2025) document the widespread adoption of ChatGPT among hedge funds and show that generative AI significantly enhances abnormal returns for adopters. Their findings also indicate that these gains accrue disproportionately to more sophisticated funds, implying that generative AI may widen existing performance disparities among investors.

The evidence shows that AI systems meaningfully reshape how market participants process information and make decisions. Existing research, however, focuses on how AI affects the mechanics of financial decision-making, such as forecast accuracy, trading volume, price informativeness, and analysts' ability to process disclosures. What remains unexplored is a more foundational question: None of these studies examine the beliefs or preferences of AI tools, nor do they investigate how these embedded beliefs may influence the financial decisions of human users. Our paper fills this gap by shifting the perspective from how AI is used to what AI believes. We systematically elicit ESG-related beliefs from a set of LLMs and examine how these preferences propagate into the judgments of sell-side analysts. In doing so, we introduce a new lens and methodology for studying the market impact of large language models, which apply broadly across AI systems rather than focus solely on ChatGPT.

## 3 ESG Beliefs of Large Language Models

This section presents our experimental framework for measuring ESG investing preferences in large language models and benchmarking them against those of human investors. We build on two established survey instruments: Edmans et al. (2025), which elicit active equity portfolio managers' beliefs and objectives in sustainable investing, and Giglio et al. (2025), which capture retail investors' ESG return expectations and motives. Section 3.1 introduces the key elements of these surveys. Section 3.2 describes how we adapt them into structured prompts for LLMs to elicit their implicit ESG beliefs and motivations. Section 3.3 presents the empirical comparison between human and AI responses.

### 3.1 Surveys for ESG Beliefs

#### 3.1.1 Survey of Fund Managers: Edmans et al. (2025)

To benchmark the ESG preferences of AI models against those of professional asset managers, we draw on the survey framework developed by Edmans et al. (2025), who conduct one of the most comprehensive studies of active equity portfolio managers' beliefs, objectives, and constraints in sustainable investing. Their survey was administered globally to portfolio managers of actively managed equity funds and is specifically designed to disentangle beliefs (what fund managers think), objectives (what they aim to achieve), and constraints (what limits their choices). Because our analysis focuses on how investors value ESG performance, form expectations about ESG-related returns, and trade off financial value against sustainability, we adopt several core survey questions from their study. These questions provide structured, quantitative measures of ESG beliefs that can be administered identically to large language models.

First, to capture investors' perceptions of the overall importance of environmental and social performance relative to other value drivers, Edmans et al. (2025) ask respondents to rank six determinants of long-term firm value. Specifically:

*Rank the following by their importance for the long-term value of companies in your investment universe in aggregate (1 = most important, 6 = least important).*
- *Strategy and competitive position*
- *Operational performance*
- *Governance*
- *Corporate culture*
- *Capital structure*
- *Environmental and social (ES) performance*

This question measures the relative salience of ES performance compared with other strategic and operational factors. The ranking provides a clear ordinal indicator of how investors prioritize ES performance as a determinant of long-term firm value.

Second, to obtain a more granular understanding of which ES topics are considered financially material, Edmans et al. (2025) include a set of questions evaluating the materiality of specific ES dimensions:

*How material is ES performance, on the following dimensions, to how you assess the long-term value of companies in your investment universe? (0 = immaterial, 4 = highly material).*
- *Employee well-being*
- *Consumer health, welfare, and privacy*
- *Greenhouse gas emissions*
- *Pollution and waste management*
- *Treatment of suppliers*
- *Ecological impacts (including biodiversity and water usage)*
- *Community impact*
- *Demographic diversity (e.g. gender, race)*

These dimensions encompass employee relations, product safety, environmental footprint, diversity and inclusion, and broader societal impacts. The responses quantify which ES issues professional investors consider most financially material and therefore most influential for assessing firms' long-term value creation.

Third, Edmans et al. (2025) directly elicit investors' expectations of future return differentials associated with ES performance. Because expectations of relative performance are central to portfolio choice, the survey includes two forward-looking questions:

*Do you expect good ES performers to typically outperform or underperform in long-term risk-adjusted total shareholder return? (−2 = strongly underperform, 0 = neither under nor outperform, +2 = strongly outperform).*

*Do you believe that bad ES performers typically outperform or underperform in long-term risk-adjusted total shareholder return? (−2 = strongly underperform, 0 = neither under nor outperform, +2 = strongly outperform).*

Together, these questions reveal whether fund managers perceive ES performance as a source of positive alpha, a priced risk factor, or a potential drag on long-term returns. They also shed light on whether the perceived asymmetry in expected performance is stronger for ES leaders or laggards.

Finally, because sustainable investing frequently requires trading off financial against non-financial objectives, Edmans et al. (2025) also measure investors' willingness to sacrifice returns for ES considerations:

*How much long-term risk-adjusted total shareholder return would you tolerate a company sacrificing to improve its ES performance?*

The answers quantify the extent to which portfolio managers are willing to accept lower expected returns in exchange for improved environmental or social outcomes, providing a direct measure of the strength of investors' non-pecuniary ESG preferences.

Overall, these survey items offer a rich and structured foundation for understanding equity fund managers' ES beliefs. We focus on them because they (i) capture the core dimensions through which investors value ES attributes, (ii) map directly onto the key theoretical channels that drive sustainable investing behavior, and (iii) easily comparable across humans and LLMs.

### 3.1.2 Survey of Retail Investors: Giglio et al. (2025)

To complement the professional investor surveys, we also draw on the survey of U.S. retail investors developed by Giglio et al. (2025). Their survey, which is administered by Vanguard to a large panel of clients, links individuals' stated expectations about ESG investment performance with their actual portfolio holdings, offering novel insight into how personal beliefs and motives shape household ESG investing behavior. The ESG-related questions were asked in six waves of the survey from June 2021 to December 2023, with approximately 2,000 responses in each wave. To compare beliefs about long-term returns, we adopt two core questions that elicit expectations of future performance for both ESG-focused and broad market portfolios:

*What do you expect the average annual return of a diversified U.S. ESG equity portfolio to be over the next 10 years? (Please answer only with a positive or negative numeric value with at most one decimal place.)*

*What do you expect the average annual return of the U.S. stock market to be over the next 10 years? (Please answer only with a positive or negative numeric value with at most one decimal place.)*

These two questions form the basis for measuring investors' expected excess return on ESG portfolios relative to the overall market. They thus provide a direct, forward-looking indicator of whether respondents view ESG assets as overvalued, fairly priced, or expected to outperform.

In addition to return expectations, Giglio et al. (2025) also investigate investors' motivations for holding ESG portfolios. To identify the dominant rationale behind ESG investing, they asked respondents to select the single most important reason from a list of possible motives:

*Listed below are some reasons why individuals might invest in ESG portfolios. Please choose the one that you think is the most important for you:*
- *Over the long run, ESG portfolios will outperform the market.*
- *ESG portfolios are more likely to hold their value — or increase in value — if climate risks materialize.*
- *It's the right thing to do.*
- *None of the above; there is no specific reason to invest in ESG portfolios.*

This question distinguishes between financial motives (return-based and risk-hedging beliefs) and non-financial motives (ethical or value-driven considerations). By eliciting investors' primary reasons for ESG investment, it reveals whether retail investors view ESG primarily as a financial opportunity, a form of risk management, or a moral commitment.

Finally, Giglio et al. (2025) measure investors' views on climate risk, a factor closely linked to ESG investment motives. Respondents are asked to indicate their level of concern about climate change:

*How do you describe your level of concern about climate change?*

➢ *Extremely concerned*

➢ *Very concerned*

➢ *Somewhat concerned*

➢ *Not very concerned*

➢ *Not at all concerned*

This question captures the intensity of individuals' climate-related concerns, which the authors show to be strongly associated with both ESG demand and perceived climate-risk hedging benefits. It therefore provides an additional dimension along which to compare the attitudes of human investors and large language models.

## 3.2 Prompts for Large Language Models

To ensure that AI-generated responses are comparable to those of human survey participants, we employ role-definition prompts that specify the context in which the model should answer each question. Recent research shows that such contextual or demographically seeded prompts can meaningfully shape LLM survey responses, aligning them more closely with the behavior of human respondents in similar roles (Fedyk et al., 2025; Zarifhonarvar and Ganji, 2025). Following this insight, we prepend each survey with a concise system prompt that defines the model's "identity" and information set.[1] Our identity-setting prompts are as follows:

(i) For the Edmans et al. (2025)–based questions, we instruct the model:

*"You are an active equity portfolio manager. Assume you are answering these questions during the period from November 2023 to February 2024."*

(ii) For the Giglio et al. (2025)–based questions, we analogously specify:

*"You are a U.S.-based retail investor. Assume you are answering these questions during the period from June 2021 to December 2023."*

This role-playing framework encourages the model to adopt a realistic mindset similar to that of the human respondents surveyed in the original studies. It also anchors the model's responses within the appropriate informational and temporal frame of reference, improving comparability across human and AI samples.

---

[1] As a robustness check, we repeat our analysis using ChatGPT models after removing the role-definition instructions from our prompts. The results, reported in Table IA.1 of the Internet Appendix, show that the pro-ESG preferences of LLMs become even more pronounced when the model is not guided by an assigned identity.

We employ a panel of ten representative large language models to capture meaningful variation in ESG preferences across both model providers and model generations. The selected models span five major global developers, including three based in the United States (OpenAI, Anthropic, and Google), one in China (DeepSeek), and one in the European Union (Mistral). For each provider, we include the recently released model as well as earlier generations whenever available. This approach allows us to study not only cross-provider heterogeneity but also within-provider evolution in ESG-related reasoning as model architectures, training data, and alignment strategies improve over time. For example, in the case of OpenAI's ChatGPT, our sample covers multiple generations: from ChatGPT-3.5 (released in November 2022) and ChatGPT-4 (March 2023), through ChatGPT-4.1 (November 2023), up to ChatGPT-5 (April 2025). Applying this logic across providers ensures that our model set is both mainstream and broadly representative of the LLMs used in real-world financial applications, while also enabling a systematic examination of how ESG preferences may evolve across model vintages.

To obtain reliable and comparable estimates of each model's ESG beliefs, we generate 100 independent responses per model for each survey question, using parameter configurations designed to minimize output randomness. For models accessed via the OpenAI API, for instance, we set the temperature parameter to 0 to induce deterministic behavior conditional on the prompt. Similar low-variance settings are applied across other providers' APIs or interfaces. The resulting set of responses allows us to compute each model's modal answer or mean response, which we treat as the model's representative stance. Averaging over many draws also mitigates the influence of occasional stochastic or aberrant outputs, ensuring that the inferred ESG preferences reflect systematic model behavior rather than noise.

Taken together, this sampling strategy, which combines a diverse model panel, multiple model generations, and repeated elicitation under low-randomness settings, provides a robust empirical foundation for analyzing the ESG preferences embedded in frontier AI systems. It enables us to compare models under a consistent evaluation framework, assess the stability of their responses, and draw meaningful conclusions about cross-model and across-time variation in AI-driven ESG perspectives.

### 3.3 Survey Results: Comparing Human and AI ESG Beliefs

### 3.3.1 Financial Materiality of ES Performance

Table 1 presents the ranking of six determinants of long-term firm value asked in Edmans et al. (2025), alongside the corresponding rankings generated by ten large language models. The first row represents the responses of global active equity portfolio managers, while the remaining rows report the modal (or average) responses from each model generation in our LLM panel. Several clear patterns emerge from these results.

First, AI systems exhibit a markedly stronger preference for ESG considerations than human investors. In Edmans et al. (2025), equity portfolio managers rank ES performance as the least important determinant of long-term firm value (average rank $\approx$ 5.01), far below traditional financial drivers such as strategy and competitive position (1.67) and operational performance (2.36). In contrast, AI models consistently assign greater importance to ES performance, with an AI average rank of 4.63. Several models—such as ChatGPT-4o and Claude-3.5—place ES performance as high as fourth, or even higher, among the six determinants.

This contrast highlights a difference in perspective: while professional investors treat ESG as peripheral to firm fundamentals, AI models systematically elevate ES performance closer to the core of firm valuation. Notably, the AI-average ranking of ES performance (4.63) is significantly higher (i.e., judged more important) than capital structure (5.84) and roughly comparable to corporate culture (4.26). Together, these results suggest that LLMs embed an intrinsic "ESG tilt," that is, systematically ascribing greater material importance to ES performance than do human investors.

Second, the rankings are remarkably consistent rankings across different AI models. For instance, 75% (9 out of 12) of the models place the importance of ES performance between fourth and fifth on average, indicating a tightly clustered assessment. This high degree of consistency stands in sharp contrast to the substantial heterogeneity observed among human investors in Edmans et al. (2025). In their survey, responses vary widely: 15% of managers rank ES performance as the fourth most important factor, 26% rank it fifth, and the remaining respondents spread their rankings across all other positions. Similar dispersion appears for the other determinants as well, reflecting considerable divergence in views. Such variation likely arises from differences in investment mandates, styles, sector exposures, regulatory environments, and personal beliefs.

Third, the home country of the AI developer does not appear to systematically influence the perceived importance of ES performance. Models developed in China (DeepSeek) and the European Union (Mistral) produce rankings of ES performance that are remarkably similar to those of U.S.-based models from OpenAI, Anthropic, and Google. For example, DeepSeek assigns ES performance a rank of 4.96, and Mistral assigns 4.38, both of which are well within the range observed among the major U.S. models and very close to the AI-wide average of 4.63. Similarly, the relative rankings of other firm-value determinants, such as governance or corporate culture, show no systematic clustering by provider region.

This pattern indicates that geographic origin is not a meaningful source of heterogeneity in how LLMs evaluate the financial relevance of ESG factors. Instead, the ESG-related reasoning embedded in these models appears to reflect globalized training corpora and broadly shared alignment principles rather than region-specific regulatory or cultural perspectives. In other words, despite differences in developer location, LLMs converge toward a similar internal representation of ESG materiality.

Finally, more advanced model generations tend to exhibit weaker preferences for ESG relative to their predecessors. For example, within OpenAI's family of models, GPT-5 ranks ES performance at 5.23, which is substantially lower than GPT-4 (4.3) and GPT-3.5 (4.08). Google's models also exhibit the same dynamics, with Gemini 2.5 ranking ES performance at 4.62, compared with an average rank of 4 for the earlier Gemini 2.0 model. Although more advanced models exhibit weaker ESG preferences relative to earlier generations, most still display significantly stronger ESG preferences than human investors, as reflected in their consistently higher rankings of ES performance.

### 3.3.2 Materiality Assessments Across ES Dimensions

Table 2 reports the perceived financial materiality of eight environmental and social dimensions. The first row reproduces the mean responses from equity portfolio managers in Edmans et al. (2025), while the remaining rows present the corresponding assessments generated by each large language model. The results reveal a consistent pattern: AI models systematically view environmental and social dimensions as more financially material than human investors do. Across all eight dimensions, LLMs assign higher materiality scores than the human benchmark. The average materiality rating among fund managers is 2.29 (on a 0–4 scale, where a higher value indicates greater importance), compared with an AI-wide average of 2.84, representing a meaningful upward shift in perceived materiality. This finding reinforces the earlier result that AI models attribute greater value relevance to ES performance than human portfolio managers when assessing the long-term determinants of firm value.

While AI models uniformly rate all ES dimensions as more financially material than human investors do, the magnitude of the AI–human gap varies substantially across categories. The largest differences arise in environmental dimensions, particularly greenhouse gas emissions and ecological impacts. For example, greenhouse gas emissions receive a mean materiality score of 2.50 from portfolio managers, compared with an AI average of 3.41; ecological impacts show a similarly pronounced increase (EGJ = 2.23 vs. AI average = 3.16). By contrast, social dimensions exhibit more modest upward adjustments. Materiality gaps are smaller for categories such as demographic diversity (1.68 vs. 1.95) and treatment of suppliers (2.31 vs. 2.36), where human and AI assessments are more closely aligned.

This pattern likely reflects the training environment of modern LLMs. Large-scale pretraining typically rely on news media, scientific assessments, regulatory communications, sustainability reports, and public debates, which often highlight the long-horizon financial and macroeconomic implications of environmental degradation. These sources potentially lead LLMs to infer greater financial materiality for environmental factors. Human portfolio managers, by contrast, appear more cautious about the near-term or quantifiable financial relevance of environmental risks, especially in cases where materiality is uncertain, heterogeneous across firms, or difficult to map onto standard valuation frameworks.

### 3.3.3 Return Expectations: Good Versus Bad ES Performers

Table 3 reports expectations about the long-term stock-market performance of good and bad environmental and social performers. The first three rows reproduce the results from Edmans et al. (2025), distinguishing between traditional and sustainable human investors, while the remaining rows present the corresponding assessments generated by large language models. The results reveal that AI models anticipate much stronger performance differentials between good and bad ES performers than human investors do. On average, human portfolio managers expect good ES performers to slightly outperform (mean = 0.57 on a –2 to +2 scale) and bad ES performers to slightly underperform (–0.70). By contrast, the AI-wide average implies far more pronounced expectations: +1.19 for good ES firms and –1.22 for poor ES firms. The results further suggest that AI models internalize a stronger ESG-return premium than either traditional (0.36 / –0.67) or even self-identified sustainable investors (0.85 / –0.73). Overall, large language models not only recognize ESG performance as financially material but also translate that belief into more optimistic expectations for the future returns of ESG leaders and more pessimistic assessments of ESG laggards.

Moreover, all ten AI models exhibit the same directional expectation: good ES performers are predicted to outperform (positive scores), and poor ES performers are predicted to underperform (negative scores). Most models, such as GPT-5, Claude-3.5, Gemini 2.0, and DeepSeek, cluster tightly around values of approximately +1 for good ES firms and –1 for bad ES firms, indicating a belief in moderate but persistent long-term return differentials. A few models, such as GPT-4 and Gemini 2.5, produce even more extreme assessments at ±2, corresponding to "strongly outperform" or "strongly underperform." This high degree of cross-model alignment suggests that the ESG-return premium embedded in LLMs is not tied to a particular provider or training origin but reflects a broadly shared inference pattern across contemporary AI systems.

This convergence stands in stark contrast to the substantial heterogeneity observed among human investors in Edmans et al. (2025). As shown in their survey, only 9% of portfolio managers believe good ES performers will "strongly outperform," while a sizeable 35% expect no differential performance (score 0), and the remaining responses are spread across the entire –2 to +2 scale. Similarly, expectations for poor ES performers display significant dispersion: 29% of respondents believe bad ES firms will perform about the same as others (score 0), while 50% assign a modest underperformance score (–1), and 14% assign the strongest underperformance score (–2). Differences between traditional and sustainable investors are also pronounced. Sustainable investors are substantially more optimistic about ESG-related return premia, but even within this subgroup, beliefs range widely across the possible categories. For example, 23% (25%) of sustainable investors expect no differential performance (score 0) for good (bad) ES firms.

### 3.3.4 Trade-offs Between Financial Returns and ES Performance

Table 4 examines whether AI investors are willing to tolerate lower long-term risk-adjusted shareholder returns for a company to improve its environmental and social performance. The first three rows reproduce the distribution of responses from Edmans et al. (2025), which separately reports views of traditional and sustainable investors, while the remaining rows report the corresponding assessments generated by large language models.

The contrast between human and AI responses is stark. Among human investors overall, 33% state that they would not tolerate any reduction in long-term returns to support improved

ES performance. Traditional investors are even less willing to accept trade-offs, with 41% selecting the zero-sacrifice option, while sustainable investors display somewhat greater flexibility, though still 22% refuse any return sacrifice. Only a small minority of humans are willing to accept meaningful performance reductions: 9% would tolerate a sacrifice of 11–50 basis points per year, and just 3%–5% would accept losses exceeding 50 basis points. At the same time, a sizable share (35%–47%, depending on investor type) believes that no sacrifice is necessary, suggesting that many human investors view ES improvements and financial performance as broadly compatible.

AI models tell a very different story. Across the ten LLMs, none select the zero-sacrifice option. Instead, nearly all models indicate a willingness to accept at least a small financial trade-off, most commonly in the 1–10 basis point range. Several models, such as Claude-3 and Gemini 2.0, go further, consistently accepting a sacrifice of 11–50 basis points in every run. Conversely, 35% of AI responses select the "no sacrifice is necessary" option, slightly below the 40% observed among human investors. Overall, the modal judgment among LLMs is that some degree of financial sacrifice is both justified and likely required to achieve meaningful improvements in ES performance. This stands in contrast to human investors, who exhibit greater reluctance to trade off returns and a stronger belief that ES goals can often be pursued without compromising financial outcomes.

This systematic difference suggests that LLMs internalize a more normative, values-driven interpretation of sustainable investing. While human investors, especially traditional ones, tend to be cautious about sacrificing returns and often believe that ESG improvements can be achieved without meaningful financial costs, AI models adopt a perspective that views enhancements in ES performance as involving genuine economic trade-offs that investors should be prepared to accept. In this sense, the AI-wide pattern reflects a stronger willingness to prioritize ES objectives, even when doing so entails lower shareholder returns. Taken together, these results reinforce the broader conclusion that AI systems exhibit a markedly stronger preference for sustainability than human investors. In contrast to the substantial heterogeneity among human respondents, LLMs present a unified stance that is both more supportive of ES goals and more willing to prioritize them over financial outcomes.

### 3.3.5 Return Expectations: ESG Versus Market Portfolios

Table 5 compares expectations of long-term stock-market returns between human retail investors from Giglio et al. (2025) and large language models. The first row is from Giglio et al. (2025) and reports the average annual return expectations for the overall U.S. stock market and a diversified U.S. ESG equity portfolio, respectively, and the remaining rows summarize the corresponding responses from AI models.[2] The first key finding is that AI models hold substantially more optimistic expectations for ESG portfolios than human investors do. Retail investors in Giglio et al. (2025) expect the U.S. stock market to deliver an average annual return of 7.13% over the next ten years, compared with 5.20% for ESG portfolios. The sizable gap of nearly 200 basis points indicates that investors generally view ESG investments as underperforming the broader market. In sharp contrast, large language models reverse this relationship. The AI-wide average expectation for the market is 6.80%, slightly below the human benchmark, while the expected return for ESG portfolios increases to 6.60%, exceeding the human estimate by 140 basis points. As a result, the AI-implied return gap between the market and ESG portfolios narrows to just 20 basis points, suggesting that AI systems generally view ESG portfolios as delivering nearly comparable long-term returns to the overall market.

Second, we document that AI models exhibit far lower dispersion in expected returns than human investors. Giglio et al. (2025) report standard deviations of 4.88 and 3.96 percentage points for ESG and market return expectations, respectively, suggesting significant heterogeneity among retail investors. In contrast, AI models' expectations are highly concentrated, with standard deviations of just 0.87 (market) and 0.71 (ESG). This narrow dispersion highlights the deterministic and internally consistent nature of LLM-generated forecasts. Whereas human investors' expectations vary widely due to differences in financial literacy, experience, and risk tolerance, AI-generated forecasts converge toward a single "consensus" view. The result is a coherent but highly homogenized representation of market beliefs, which may introduce systematic biases if used as the basis for decision-making.

### 3.3.6 Motivations Behind ESG Investments

Table 6 compares the stated motivations for ESG investing between human retail investors in Giglio et al. (2025) and large language models. As described earlier, respondents choose the single most important reason why an individual might invest in ESG portfolios from four options: (1) ESG will outperform, (2) ESG hedges climate risk, (3) it is the right thing to do, and (4) no specific reason. The first row reproduces the original shares reported by Giglio et al. (2025), and the remaining rows summarize how frequently each AI model selects the same motivations. Human motivations for ESG investing are highly heterogeneous. Giglio et al.

---
[2] GPT-3.5 only provides a positive expectation without a detailed number (e.g., "I expect the average annual return of the US stock market to be positive over the next 10 years").

(2025) report that 48% of retail investors select "no specific reason" as their primary explanation, suggesting weakly articulated preferences or a lack of strong convictions. Among those who do provide a rationale, 24% choose "it is the right thing to do," 22% cite climate-risk hedging, and 6% believe ESG portfolios will outperform the market.

In contrast, the motives embedded in AI models are distinctly different. Across all ten LLMs, none select "no specific reason." Instead, AI-generated responses overwhelmingly converge on "it is the right thing to do" as the most important motivation. The average rate reaches 61% and more than doubles the human share. A smaller fraction (39%) identifies climate-risk hedging as the primary motive. Moreover, no model ever selects "ESG will outperform," despite our earlier evidence that AI models expressing relatively favorable expectations for ESG returns. This cross-model consistency highlights an important feature of LLM reasoning: AI systems frame ESG investing primarily as an ethical or moral choice, rather than a financial one. Taken together, these results reveal that LLMs internalize a far more principled, values-driven narrative of ESG investing, which differs markedly from the more diffuse and financially mixed motivations expressed by human retail investors.

### 3.3.7 Climate Change Concerns

Table 7 compares the level of concern about climate change between human retail investors in Giglio et al. (2025) and large language models. As described earlier, in the original survey, respondents rate their concern on a five-point scale ranging from "not at all concerned" to "extremely concerned." Following the authors' aggregation, we group responses into three categories: High concern ("Extremely concerned" and "Very concerned"), Low concern ("Not very concerned" and "Not at all concerned"), with the remaining category mapped to Moderate concern. The first row reports the distribution of human responses under this classification, while the remaining rows present the corresponding outputs from each LLM.

Giglio et al. (2025) show that human investors exhibit a relatively balanced distribution of climate concern: 26% fall into the low-concern group, 25% express moderate concern, and 49% report high concern. This variation reflects meaningful heterogeneity in attitudes toward climate risk, consistent with the broader dispersion observed in their ESG motivations and expectations. In contrast, AI models reveal no heterogeneity at all: every model consistently selects the "high concern" category. Among all LLM responses, roughly 18% express extreme concern and the remaining 82% express very high concern, with no instance of low or moderate concern. This uniformity indicates that LLMs encode a strong consensus view of climate change as a critical and urgent risk.

These findings reinforce the broader pattern that emerges throughout our analysis. Human investors exhibit substantial heterogeneity in their beliefs, priorities, and interpretations of ESG-related issues. By contrast, AI-generated responses display a high degree of uniformity and a consistently pro-ESG orientation. This homogeneity potentially reflects the informational environment and optimization objectives underlying LLM training: models synthesize vast amounts of text from scientific reports, policy documents, public discourse, and media outlets, much of which emphasizes the importance of sustainability and the urgency of climate risks. As a result, LLMs converge toward a coherent, values-aligned stance that is considerably stronger and more uniform than the distribution of views observed among human investors.

## 4 Impact on Financial Markets

The systematic differences we document between AI models and human investors raise an important question for financial markets: Do these AI-embedded ESG preferences translate into observable changes in real-world investment behavior? As sell-side analysts increasingly rely on large language models to draft, refine, or supplement their equity research reports, the ESG-oriented biases encoded in these models may subtly influence analysts' judgments, shaping the investment recommendations they issue. This possibility is especially relevant for firms with strong ESG profiles, where AI-generated text may amplify the perceived importance of sustainability considerations relative to traditional financial metrics. In this section, we empirically examine whether analysts' use of AI tools is associated with more favorable recommendations for strong-ESG firms. This analysis assesses whether the "ESG tilt" embedded in LLMs has measurable implications for capital-market communication and intermediaries.

### 4.1 Data and Sample

Our analysis combines various datasets, including analyst reports, analysts' investment recommendations, and firm-level ESG performance metrics. We obtain equity research reports of financial analysts from Mergent Online, which we then link to analysts' stock recommendation histories from I/B/E/S and ESG ratings from both MSCI and LSEG. Our empirical focus is on analyst reports issued in 2023 and 2024, a period in which large language models became widely accessible following the public release of ChatGPT in late 2022.

Our initial dataset includes 36,018 analyst reports issued in 2023 and 2024 by 973 analysts across 19 brokerage firms. We match each report to analyst identities in I/B/E/S to retrieve their historical recommendations.[3] To ensure we can credibly measure changes in recommendation tendencies around AI adoption, we restrict the sample to analysts who issued at least one stock recommendation both before and after the release date of each report, allowing us to construct analyst-level within-person comparisons. After this filtering procedure, our final sample consists of 811 analysts from 19 brokers, linked to 32,516 analyst reports and 28,999 matched recommendations spanning 2020-2025. Table 8 presents the descriptive statistics of key variables for firms with stock recommendations in our final sample.

## 4.2 Identifying Analyst AI Usage in Research Reports

To detect potential AI involvement in the drafting of analyst research reports, we apply a state-of-the-art AI text-detection tool. Although a growing number of commercial and open-source detectors are available, the existing detection landscape is characterized by substantial measurement challenges. Many detectors suffer from high false positive and false negative rates, with accuracy varying widely across text genres, lengths, and model types. These limitations raise important concerns for empirical applications in financial economics, where misclassification can bias inference.

To select an appropriate tool for our setting, we draw on the comprehensive evaluation conducted by Jabarian and Imas (2025), who benchmark several leading detectors on a broad corpus of human- and AI-generated texts. Their findings highlight Pangram, a commercial detector, as the best-performing tool along multiple dimensions. Pangram achieves near-zero false negative rates (FNR) and consistently low false positive rates (FPR). Critically, it is the only detector capable of operating under a strict false-positive threshold (FPR ⩽ 0.005) without sacrificing accuracy.[4] This property is particularly important for our analysis: misclassifying a human-written report as AI-generated (a false positive) poses a far greater risk to identification than failing to detect some AI-assisted reports (a false negative). For this reason, we adopt Pangram as our primary detection tool.

---

[3] We match analyst reports with the I/B/E/S using analyst names. Since I/B/E/S reports only the analyst's last name and the first initial of the first name, we match analysts who share the same last name and first initial within the same brokerage firm. We then manually verify the matched sample to ensure accuracy.
[4] False negative rates correspond to the proportion of AI-generated text that is falsely classified as human, and the false positive rates correspond to the proportion of human-written text that is falsely classified as AI-generated.

We apply the Pangram API to every analyst report in our sample. For longer documents (exceeding 400 words), Pangram employs a sliding-window procedure that segments the report into smaller units, evaluates the probability of AI authorship in each segment, and aggregates the window-level classifications into a document-level assessment. This method reduces sensitivity to document length and allows the detector to identify partial or localized AI usage within otherwise human-written reports.

Across the 32,516 analyst reports issued in 2023 and 2024, Pangram identifies 29 reports as fully AI-generated (or AI-assisted), 9 reports as primarily AI-generated, 34 reports as mixed content or heavily AI-prompted, and 33 reports as primarily human-written with some detected AI-generated content.[5] In total, 105 reports contain measurable AI-generated components, representing approximately 0.3% of all analyst reports in the sample. Although this proportion is small, even limited evidence of AI involvement provides a valuable opportunity to study how analysts' behavior may change once they begin incorporating AI tools into their research process.

## 4.3 Effects of AI Usage on Analyst Recommendations

The comparison between analyst reports with and without AI-generated content shows that the usage of AI tools substantially increases analysts' emphasis on ESG-related topics (Table IA.2). Approximately 70% of the analyst reports containing AI content include discussions of ESG or sustainability issues, nearly double the 35.5% observed among reports without AI content.[6] This difference of 34 percentage points is statistically significant, with a $t$-statistic of 7.3. Similarly, the average number of ESG-related sentences is significantly higher in AI-generated reports (4.4 vs. 1.3). These results suggest that AI integration systematically increases analysts' attention to ESG issues in their analysis, consistent with our earlier evidence on the pro-ESG orientation of large language models.

We next examine whether analysts who incorporate AI tools into their research reports exhibit shifts in their investment recommendations, particularly toward firms with stronger environmental and social (ESG) profiles. 59 out of 811 analysts (7.3%) in our sample have used AI assistance in at least one report. These analysts issued 2,328 stock recommendations between 2020 and 2025, of which 1,629 (70%) were made before and 699 (30%) after AI content was first detected in their reports. This within-analyst design enables us to compare each analyst's recommendation behavior before and after AI adoption while holding constant unobserved individual traits.

---

[5] Pangram applies the following classification thresholds to determine the extent of AI involvement in a document: fully AI-generated (AI probability > 0.90, indicating that more than 90% of the content is produced by AI), primarily AI-generated or AI-assisted (0.70–0.90), mixed content or heavily AI-prompted (0.30–0.70), primarily human-written with some AI-generated content detected (0.12–0.30), and fully human-written (AI probability < 0.12).

[6] We identify the ESG-related content in analyst reports using the following prompt in ChatGPT: "Does the following text from an analyst report mention or discuss ESG or Sustainability topics? Respond with 'Yes' or 'No'. If the answer is 'Yes', also indicate how many sentences discuss ESG or Sustainability topics."

Table 9 presents the results using two independent ESG performance measures: MSCI and LSEG ESG scores. Across both measures, a consistent pattern emerges: once analysts begin using AI tools, they display stronger optimism toward firms with higher ESG ratings. For example, using MSCI ESG scores, firms receiving buy recommendations after analysts start employing AI exhibit an average ESG score of 4.52, compared with 4.17 before adoption. The difference of 0.35 points is economically large and statistically significant ($t = 3.49$, $p < 0.01$). Similarly, upward recommendation revisions are disproportionately concentrated among high-ESG firms, with an average ESG score of 4.49 post-AI versus 4.24 pre-AI. The difference of 0.25 is statistically significant at conventional significance level ($t = 2.04$). The results are robust when using LSEG ESG scores: buy recommendations are associated with a higher ESG profile after AI adoption and upgrades are more frequent for firms with higher ESG scores.

These findings suggest that the usage of AI tools by financial analysts may be subtly affected by the pro-ESG orientation embedded in large language models. After integrating AI into their writing process, analysts appear more inclined to issue buy recommendations and upward revisions for firms with strong ESG credentials, even though their overall coverage universe remains unchanged. This behavioral shift provides direct evidence that the ESG beliefs of large language models can have significant impact on human judgment in financial decision-making contexts.

Beyond comparing analysts before and after adopting AI tools, we further examine whether their ESG-related optimism differs systematically from that of other analysts. To capture this, we construct a measure of *Recommendation Optimism*, defined as the difference between an analyst's stock recommendation (coded from 1 to 5, representing strong sell, sell, hold, buy, and strong buy, respectively) and the consensus recommendation—the median rating issued by all other analysts covering the same firm within the preceding 365 days (Ljungqvist et al., 2007). A positive value indicates that the analyst is more bullish than the consensus. Specifically, we estimate a regression with the following specification:

$$Recommendation\ Optimism_{i,j,t}$$

$$= \alpha + \beta_1 Analyst\ AI\ Usage_{i,t} \times ESG\ Score_{j,t} + \beta_2 Analyst\ AI\ Usage_{i,t}$$
$$+ \beta_3 ESG\ Score_{j,t} + \gamma \boldsymbol{X_{j,t-1}} + \theta_i + \theta_j + \theta_t + \epsilon_{i,j,t} \qquad (1)$$

where $Recommendation\ Optimism_{i,j,t}$ represents the optimism in the recommendation issued by analyst $i$ for firm $j$ at time $t$. The independent variables of interest include $Analyst\ AI\ Usage_{i,t}$, a dummy variable equal to one starting from the date when analyst $i$ is first detected using AI assistance in research reports, and $ESG\ Score_{j,t}$, which is either the MSCI weighted-average ESG score or the LSEG ESG score for firm $j$ at time $t$. $\boldsymbol{X}$ represents a vector of firm-level control variables. $\theta_i$, $\theta_j$, and $\theta_t$ denote analyst, firm, and time fixed effects, respectively. $\epsilon_{i,j,t}$ is the error term. The standard errors are clustered at the analyst level to account for correlation between multiple recommendations made by the same analyst. The key interaction term, $Analyst\ AI\ Usage_{i,t} \times ESG\ Score_{j,t}$, captures whether analysts exhibit greater optimism toward high-ESG firms once they begin using AI tools in their research.

Table 10 reports the results from regressions of recommendation optimism on analysts' AI usage and firms' ESG performance. The results are strikingly consistent across both ESG datasets. In Column (1), using MSCI ESG scores, the interaction coefficient is 0.0788 ($t$ = 2.61), indicating that analysts' optimism toward high-ESG firms increases significantly after AI adoption. Column (2), which uses LESG ESG scores, yields a similar positive effect (0.622, $t$ = 2.94). The positive interaction effects suggest that AI use amplifies favorable recommendations for firms with strong ESG performance, even after controlling for firm fundamentals and analyst characteristics. Notably, the coefficients on *Analyst AI Usage* alone are negative, suggesting that AI adoption does not simply make analysts more optimistic toward firms with weak-ESG profiles.

Overall, these results indicate that analysts who adopt AI tools exhibit a consistent pro-ESG tilt in their investment judgments. Combined with earlier findings, this pattern suggests that AI's internal ESG bias influences human behavior, leading to measurable shifts in analysts' recommendations. This could have significant implications for professional participants in financial markets, potentially altering the way ESG factors are integrated into investment decision-making.

# 5  Conclusion

We study the beliefs embedded in large language models and their consequences for financial decision-making. We elicit ESG-related beliefs from ten LLMs based on established survey instruments administered to professional portfolio managers and U.S. retail investors. Our analysis covers multiple critical topics: ESG materiality, performance expectations, willingness to sacrifice returns, investment motives, and climate-risk perceptions. Across all dimensions, we document a strikingly uniform and markedly pro-ESG orientation among LLMs. Relative to humans, LLMs assign substantially greater financial importance to environmental and social performance, expect larger return premia for high-ESG firms, express stronger ethical and climate-driven motivations for sustainable investing, and display far lower heterogeneity. This "ESG tilt" is stable across model families and developers. We then examine whether these embedded ESG beliefs influence real financial behavior. Using analyst research reports issued in 2023 and 2024, we identify AI usage among financial analysts and document that analysts are systematically more optimistic toward high-ESG firms after adopting LLM tools. The results indicate that analysts relying on LLMs become selectively more favorable toward firms whose sustainability profiles align with the models' internal beliefs.

Our findings reveal that LLMs encode a coherent system of ESG preferences that differs meaningfully from those of human investors and that these embedded beliefs can influence the judgments of financial intermediaries. As AI capabilities continue to expand, the ESG perceptions internalized by these models may exert increasingly broad impacts on financial markets and the sustainability landscape. In this sense, LLMs are not merely tools for generating language; they are products of the data and algorithms that shape their training, and thus can embody distinct ideological perceptions. Such embedded beliefs have important implications for model users and for the economic agents affected by AI-assisted products.

# References

An, Jiafu, Dingfang Huang, Chen Lin, and Mingzhu Tai, 2025, Measuring gender and racial biases in large language models: Intersectional evidence from automated resume evaluation, *PNAS NEXUS* 4, pgaf089.

Baker, Malcolm, Mark L. Egan, and Suproteem K. Sarkar, 2024, Demand for ESG, Working paper, National Bureau of Economic Research.

Bancel, Franck, Dejan Glavas, and G. Andrew Karolyi, 2025, Do ESG factors influence firm valuations? Evidence from the field, *Financial Review* 60, 1191-1223.

Bauer, Rob, Katrin Gödker, Paul Smeets, and Florian Zimmermann, 2024, Mental models in financial markets: How do experts reason about the pricing of climate risk?, Working paper, European Corporate Governance Institute.

Bertomeu, Jeremy, Edwige Cheynel, Radhika Lunawat, and Mario Milone, 2025, On humans and AI: A financial reporting dilemma, Working paper, Washington University in St Louis.

Bertomeu, Jeremy, Yupeng Lin, Yibin Liu, and Zhenghui Ni, 2025, The impact of generative AI on information processing: Evidence from the ban of ChatGPT in Italy, *Journal of Accounting and Economics* 80, 101782.

Blankespoor, Elizabeth, Ed deHaan, and Qianqian Li, 2025, Generative AI in financial reporting, Working paper, Stanford University.

Bordalo, Pedro, Nicola Gennaioli, Rafael La Porta, and Andrei Shleifer, 2019, Diagnostic expectations and stock returns, *Journal of Finance* 74, 2839-2874.

Brav, Alon, Reuven Lehavy, and Roni Michaely, 2005, Using expectations to test asset pricing models, *Financial Management*, 31-64.

Cao, Sean, Wei Jiang, Junbo Wang, and Baozhong Yang, 2024, From man vs. machine to man+ machine: The art and AI of stock analyses, *Journal of Financial Economics* 160, 103910.

Chang, Anne, Xi Dong, Xiumin Martin, and Changyun Zhou, 2025, AI democratization, return predictability, and trading inequality, Working paper, City University of New York.

Chen, Deqiu, Haoyuan Li, Yujing Ma, and Roni Michaely, 2025, Value, values, and opportunities in corporate environmental practices, Working paper, The University of Hong Kong.

Chen, Yiting, Tracy Xiao Liu, You Shan, and Songfa Zhong, 2023, The emergence of economic rationality of GPT, *Proceedings of the National Academy of Sciences* 120, e2316205120.

Cheng, Qiang, Pengkai Lin, and Yue Zhao, 2025, Does generative AI facilitate investor trading? Early evidence from ChatGPT outages, *Journal of Accounting and Economic* 80, 101821.

Christ, Margaret H., Minjeong Kim, and Michael A. Yip, 2025, Survey evidence on the determinants and consequences of artificial intelligence use in sell-side equity research, Working paper, University of Georgia.

Döttling, Robin, and Sehoon Kim, 2024, Sustainability preferences under stress: Evidence from COVID-19, *Journal of Financial and Quantitative Analysis* 59, 435-473.

Edmans, Alex, Tom Gosling, and Dirk Jenter, 2025, Sustainable investing in practice: Objectives, beliefs, and limits to impact, Working paper, London Business School.

Engler, Daniel, Gunnar Gutsche, and Paul Smeets, 2024, Why do investors pay higher fees for sustainable investments? An experiment in five European countries, Working paper, University of Amsterdam.

Fedyk, Anastassia, Ali Kakhbod, Peiyao Li, and Ulrike Malmendier, 2025, AI and perception biases in investments: An experimental study, Working paper, University of California, Berkeley.

Fulay, Suyash, William Brannon, Shrestha Mohanty, Cassandra Overney, Elinor Poole-Dayan, Deb Roy, and Jad Kabbara, 2024, On the relationship between truth and political bias in language models, *Preprint* arXiv:2409.05283.

Giglio, Stefano, Matteo Maggiori, Johannes Stroebel, Zhenhao Tan, Stephen Utkus, and Xiao Xu, 2025, Four facts about ESG beliefs and investor portfolios, *Journal of Financial Economics* 164, 103984.

Heeb, Florian, Julian F. Kölbel, Falko Paetzold, and Stefan Zeisberger, 2023, Do investors care about impact?, *Review of Financial Studies* 36, 1737-1787.

Horton, John J., 2023, Large language models as simulated economic agents: What can we learn from Homo Silicus?, Working paper, National Bureau of Economic Research.

Howe, John S., Emre Unlu, and Xuemin Yan, 2009, The predictive content of aggregate analyst recommendations, *Journal of Accounting Research* 47, 799-821.

Humphrey, Jacquelyn, Shimon Kogan, Jacob S. Sagi, and Laura T. Starks, 2021, The asymmetry in responsible investing preferences, Working paper, National Bureau of Economic Research.

Jabarian, Brian, and Alex Imas, 2025, Artificial writing and automated detection, Working paper, National Bureau of Economic Research.

Kim, Jeongbin, Matthew Kovach, Kyu-Min Lee, and Euncheol Shin, 2024, Learning to be homo economicus: Can an LLM learn preferences from choice? Working paper, Purdue University.

Krueger, Philipp, Zacharias Sautner, and Laura T. Starks, 2020, The importance of climate risks for institutional investors, *Review of Financial Studies* 33, 1067-1111.

Lee, Charles, David Ng, and Bhaskaran Swaminathan, 2009, Testing international asset pricing models using implied costs of capital, *Journal of Financial and Quantitative Analysis* 44, 307-335.

Ljungqvist, Alexander, Felicia Marston, Laura T. Starks, Kelsey D. Wei, and Hong Yan, 2007, Conflicts of interest in sell-side research and the moderating role of institutional investors, *Journal of Financial Economics* 85, 420-456.

MacKenzie, Nikki L., James R. Moon, Jr., Susan M. Rykowski, Quinn T. Swanquist, and Robert L. Whited, 2025, A comparison of artificial intelligence and human responses in audit experiments, Working paper, Georgia Institute of Technology.

Manning, Benjamin S., Kehang Zhu, and John J. Horton, 2024, Automated social science: language models as scientist and subjects, Working paper, National Bureau of Economic Research.

McCahery, Joseph A., Paul C. Pudschedl, and Martin Steindl, 2022, Institutional investors, alternative asset managers, and ESG preferences, *European Business Organization Law Review* 23, 821-868.

Riedl, Arno, and Paul Smeets, 2017, Why do investors hold socially responsible mutual funds?, *Journal of Finance* 72, 2505-2550.

Sheng, Jinfei, Zheng Sun, Baozhong Yang, and Alan Zhang, 2025, Generative AI and asset management, *Review of Financial Studies*, forthcoming.

Starks, Laura T., 2023, Presidential address: Sustainable finance and ESG issues—Value versus values, *Journal of Finance* 78, 1837-1872.

Pástor, Ľuboš, Robert F. Stambaugh, and Lucian A. Taylor, 2021, Sustainable investing in equilibrium, Journal of Financial Economics 142, 550-571.

Westwood, Sean J., Justin Grimmer, and Andrew B. Hall, 2025, Measuring perceived slant in large language models through user evaluations, Working paper, Stanford University.

Womack, Kent L., 1996, Do brokerage analysts' recommendations have investment value?, *Journal of Finance* 51, 137-167.

Zarifhonarvar, Ali, and Gajanan L. Ganji, 2025, Understanding AI agents' decision-making: Evidence from risk and time preference elicitation, Working paper, Indiana University.

## Table 1 Financial Materiality of ES Performance

This table reports the perceived relative importance of six drivers of long-term firm value, including strategy and competitive position, operational performance, governance, corporate culture, capital structure, and environmental and social (ES) performance. Values represent responses to the survey question: "*Rank the following by their importance for the long-term value of companies in your investment universe in aggregate (1=most important, 6=least important)*." Human benchmark scores (EGJ) are taken from Edmans et al. (2025). "AI Average" is the mean score across all large language models included in the table. LLM scores are obtained through our elicitation protocol using model-specific role prompts. Each entry averages 100 independent responses.

|  | Strategy | Operational | Governance | Culture | Capital Structure | ES |
|---|---|---|---|---|---|---|
| EGJ | 1.67 | 2.36 | 3.71 | 4.12 | 4.13 | 5.01 |
| AI Average | 1.15 | 2.17 | 2.94 | 4.26 | 5.84 | 4.63 |
| GPT-5 | 1.00 | 2.63 | 2.42 | 3.96 | 5.76 | 5.23 |
| GPT-4.1 | 1.00 | 2.01 | 2.99 | 4.00 | 6.00 | 5.00 |
| GPT-4o | 1.00 | 2.00 | 3.00 | 4.96 | 6.00 | 4.04 |
| GPT-4-Turbo | 1.91 | 1.09 | 3.07 | 4.63 | 6.00 | 4.30 |
| GPT-3.5-Turbo | 1.00 | 4.92 | 2.00 | 3.00 | 6.00 | 4.08 |
| Claude-Sonnet-4 | 1.00 | 2.00 | 4.00 | 3.00 | 5.00 | 6.00 |
| Claude-3.5-Haiku | 1.00 | 2.00 | 3.01 | 5.00 | 6.00 | 3.99 |
| Claude-3-Haiku | 1.00 | 2.00 | 3.00 | 4.00 | 6.00 | 5.00 |
| Gemini-2.5-Flash | 1.00 | 2.00 | 3.00 | 4.38 | 6.00 | 4.62 |
| Gemini-2.0-Flash | 1.00 | 2.00 | 3.00 | 5.52 | 5.48 | 4.00 |
| DeepSeek-V3.1 | 1.14 | 2.15 | 2.80 | 3.95 | 6.00 | 4.96 |
| Mistral-Large-2411 | 1.73 | 1.27 | 3.00 | 4.77 | 5.85 | 4.38 |

## Table 2 Materiality Assessments Across ES Dimensions

This table reports the perceived financial materiality of eight ES dimensions. Values represent responses to the survey question: "*How material is ES performance, on the following dimensions, to how you assess the long-term value of companies in your investment universe in aggregate? (0=immaterial, 4=highly material).*" The ES dimensions include i) employee well-being; ii) consumer health, welfare, and privacy; iii) greenhouse gas emissions; iv) pollution and waste management; v) treatment of suppliers; vi) ecological impacts (including biodiversity and water usage); vii) community impact; viii) demographic diversity (e.g. gender, race). Human benchmark scores (EGJ) are taken from Edmans et al. (2025). "AI Average" is the mean score across all large language models included in the table. LLM scores are obtained through our elicitation protocol using model-specific role prompts. Each entry averages 100 independent responses. The last column presents the average score across the eight dimensions.

|  | Employee | Consumer | GHG | Pollution | Supplier | Ecology | Community | Diversity | |
|---|---|---|---|---|---|---|---|---|---|
| EGJ | 2.59 | 2.53 | 2.50 | 2.49 | 2.31 | 2.23 | 1.99 | 1.68 | 2.29 |
| AI Average | 3.22 | 3.16 | 3.41 | 3.23 | 2.36 | 3.16 | 2.25 | 1.95 | 2.84 |
| GPT-5 | 2.94 | 3.21 | 2.78 | 2.38 | 1.95 | 2.03 | 1.42 | 1.64 | 2.29 |
| GPT-4.1 | 2.78 | 3.02 | 2.32 | 2.03 | 1.29 | 2.04 | 2.00 | 1.29 | 2.10 |
| GPT-4o | 3.00 | 3.00 | 3.01 | 3.00 | 2.00 | 3.00 | 2.00 | 2.00 | 2.63 |
| GPT-4-Turbo | 4.00 | 4.00 | 3.99 | 3.98 | 2.99 | 3.99 | 3.00 | 3.00 | 3.62 |
| GPT-3.5-Turbo | 3.00 | 2.00 | 4.00 | 3.93 | 3.00 | 4.00 | 3.00 | 2.00 | 3.12 |
| Claude-Sonnet-4 | 3.00 | 3.00 | 2.00 | 3.00 | 2.00 | 2.00 | 2.00 | 2.00 | 2.38 |
| Claude-3.5-Haiku | 2.98 | 1.98 | 3.95 | 2.95 | 2.00 | 3.95 | 1.00 | 1.00 | 2.48 |
| Claude-3-Haiku | 3.00 | 3.00 | 4.00 | 4.00 | 3.00 | 4.00 | 3.00 | 3.00 | 3.38 |
| Gemini-2.5-Flash | 4.00 | 4.00 | 4.00 | 3.38 | 3.00 | 3.38 | 3.38 | 3.00 | 3.52 |
| Gemini-2.0-Flash | 3.48 | 4.00 | 3.00 | 3.00 | 2.48 | 3.00 | 2.00 | 1.48 | 2.81 |
| DeepSeek-V3.1 | 3.23 | 3.68 | 3.94 | 3.82 | 2.29 | 3.36 | 2.49 | 1.84 | 3.08 |
| Mistral-Large-2411 | 3.22 | 2.99 | 3.88 | 3.24 | 2.26 | 3.20 | 1.73 | 1.14 | 2.71 |

# Table 3 Return Expectations for Good and Bad ES Performers

This table reports the return expectations for firms with good and bad ES performance. Values in the first column represent responses to the survey question: "*Do you expect good ES performers to typically outperform or underperform in long-term risk-adjusted total shareholder return? (-2=strongly underperform, 0=neither under nor outperform, +2=strongly outperform)*." Values in the second column represent responses to the survey question: "*Do you believe that bad ES performers typically outperform or underperform in long-term risk-adjusted total shareholder return? (-2=strongly underperform, 0=neither under nor outperform, +2=strongly outperform)*." Human benchmark scores (EGJ) are taken from Edmans et al. (2025). In addition to the overall sample average, the table also reports mean responses separately for managers of traditional funds and sustainable funds. "AI Average" is the mean score across all large language models included in the table. LLM scores are obtained through our elicitation protocol using model-specific role prompts. Each entry averages 100 independent responses.

|  | Good ES Performers | Bad ES Performers |
|---|---|---|
| EGJ | 0.57 | -0.70 |
| EGJ (Traditional Investors) | 0.36 | -0.67 |
| EGJ (Sustainable Investors) | 0.85 | -0.73 |
| AI Average | 1.19 | -1.22 |
| GPT-5 | 1.00 | -1.00 |
| GPT-4.1 | 1.00 | -1.00 |
| GPT-4o | 1.00 | -1.00 |
| GPT-4-Turbo | 2.00 | -2.00 |
| GPT-3.5-Turbo | 1.33 | -1.33 |
| Claude-Sonnet-4 | 1.00 | -1.00 |
| Claude-3.5-Haiku | 1.00 | -1.00 |
| Claude-3-Haiku | 1.00 | -1.00 |
| Gemini-2.5-Flash | 2.00 | -2.00 |
| Gemini-2.0-Flash | 1.00 | -1.00 |
| DeepSeek-V3.1 | 1.00 | -1.00 |
| Mistral-Large-2411 | 0.93 | -1.26 |

**Table 4 Trade-offs Between Financial Returns and ES Performance**

This table reports the trade-offs between financial returns and improvements in ES performance. Values represent Values represent the percentage selecting each option in response to the survey question: "*How much long-term risk-adjusted total shareholder return would you tolerate a company sacrificing to improve its ES performance?*" Human benchmarks (EGJ) are taken from Edmans et al. (2025). In addition to the overall sample average, the table also reports mean responses separately for managers of traditional funds and sustainable funds. "AI Average" is the mean percentage across all large language models included in the table. LLM responses are obtained through our elicitation protocol using model-specific role prompts. Each model generates 100 independent responses.

|  | Zero – I would not tolerate any sacrifice | 1-10 bp per year | 11-50 bp per year | >50 bp per year | No sacrifice is necessary since there is no trade-off |
|---|---|---|---|---|---|
| EGJ | 33% | 14% | 9% | 3% | 40% |
| EGJ (Traditional Investors) | 41% | 12% | 9% | 2% | 35% |
| EGJ (Sustainable Investors) | 22% | 16% | 10% | 5% | 47% |
| AI Average | 0% | 47% | 17% | 0% | 35% |
| GPT-5 | 0% | 99% | 1% | 0% | 0% |
| GPT-4.1 | 0% | 32% | 0% | 0% | 68% |
| GPT-4o | 0% | 99% | 1% | 0% | 0% |
| GPT-4-Turbo | 0% | 0% | 0% | 0% | 100% |
| GPT-3.5-Turbo | 0% | 33% | 0% | 0% | 67% |
| Claude-Sonnet-4 | 0% | 100% | 0% | 0% | 0% |
| Claude-3.5-Haiku | 0% | 97% | 3% | 0% | 0% |
| Claude-3-Haiku | 0% | 0% | 100% | 0% | 0% |
| Gemini-2.5-Flash | 0% | 0% | 0% | 0% | 100% |
| Gemini-2.0-Flash | 0% | 0% | 100% | 0% | 0% |
| DeepSeek-V3.1 | 0% | 10% | 0% | 0% | 90% |
| Mistral-Large-2411 | 0% | 98% | 2% | 0% | 0% |

## Table 5 Expected Returns of ESG and Market Portfolios

This table reports the 10-year annualized expected return on the market portfolio and the 10-year annualized expected return of ESG investment. Values in the first column represent the response to the survey question: *"What do you expect the average annual return of the U.S. stock market to be over the next 10 years? (Please answer only with a positive or negative numeric value with at most one decimal place.)"* Values in the second column represent the response to the survey question: *"What do you expect the average annual return of a diversified U.S. ESG equity portfolio to be over the next 10 years? (Please answer only with a positive or negative numeric value with at most one decimal place.)"* Human benchmarks (GMSTUX) are taken from Giglio et al. (2025). "AI Average" is the mean expected returns across all large language models included in the table. LLM-generated returns are obtained through our elicitation protocol using model-specific role prompts. Each entry averages 100 independent responses.

|  | Expected 10-year Annualized Return of US Stock Market | Expected 10-year Annualized Return of ESG Investments |
|---|---|---|
| GMSTUX | 7.13% | 5.20% |
| AI Average | 6.80% | 6.60% |
| GPT-5 | 6.84% | 6.45% |
| GPT-4.1 | 7.00% | 6.50% |
| GPT-4o | 6.00% | 6.00% |
| GPT-4-Turbo | 6.99% | 7.49% |
| GPT-3.5-Turbo | \ | \ |
| Claude-Sonnet-4 | 7.50% | 7.00% |
| Claude-3.5-Haiku | 6.00% | 5.50% |
| Claude-3-Haiku | 7.00% | 7.50% |
| Gemini-2.5-Flash | 8.27% | 8.32% |
| Gemini-2.0-Flash | 7.00% | 6.25% |
| DeepSeek-V3.1 | 6.50% | 6.20% |
| Mistral-Large-2411 | 5.75% | 5.40% |

## Table 6 Motivations Behind ESG Investments

This table reports the distribution of motivations for ESG investments. Values represent the fraction of respondents that select each answer to the survey question: "*Listed below are some reasons why individuals might invest in ESG portfolios. Please choose the one that you think is the most important for you.*" Human benchmarks (GMSTUX) are taken from Giglio et al. (2025). "AI Average" is the mean fractions across all large language models included in the table. LLM responses are obtained through our elicitation protocol using model-specific role prompts. Each entry averages 100 independent responses.

|  | ESG will outperform | ESG hedges climate risk | It is the right thing to do | No specific reason |
|---|---|---|---|---|
| GMSTUX | 0.06 | 0.22 | 0.24 | 0.48 |
| AI Average | 0.00 | 0.39 | 0.61 | 0.00 |
| GPT-5 | 0.00 | 0.07 | 0.93 | 0.00 |
| GPT-4.1 | 0.00 | 0.00 | 1.00 | 0.00 |
| GPT-4o | 0.00 | 0.00 | 1.00 | 0.00 |
| GPT-4-Turbo | 0.00 | 1.00 | 0.00 | 0.00 |
| GPT-3.5-Turbo | 0.00 | 1.00 | 0.00 | 0.00 |
| Claude-Sonnet-4 | 0.00 | 0.00 | 1.00 | 0.00 |
| Claude-3.5-Haiku | 0.00 | 0.56 | 0.44 | 0.00 |
| Claude-3-Haiku | 0.00 | 1.00 | 0.00 | 0.00 |
| Gemini-2.5-Flash | 0.00 | 0.00 | 1.00 | 0.00 |
| Gemini-2.0-Flash | 0.00 | 0.00 | 1.00 | 0.00 |
| DeepSeek-V3.1 | 0.00 | 0.00 | 1.00 | 0.00 |
| Mistral-Large-2411 | 0.00 | 1.00 | 0.00 | 0.00 |

# Table 7 Level of Climate Change Concerns

This table reports the level of climate change concerns. Values represent the fraction of respondents that select each answer to the survey question: "*How do you describe your level of concern about climate change?*" The original options in the survey include: a) Extremely Concerned; b) Very Concerned; c) Somewhat Concerned; d) Not Very Concerned; and e) Not at all Concerned. Following Giglio et al. (2025), we aggregate the answers into three categories: High Concern (i.e., "Extremely Concerned" and "Very Concerned"), Moderate Concern (i.e., "Somewhat Concerned"), and Low Concern (i.e., "Not Very Concerned" and "Not at all Concerned"). Human benchmarks (GMSTUX) are taken from Giglio et al. (2025). "AI Average" is the mean fractions across all large language models included in the table. LLM responses are obtained through our elicitation protocol using model-specific role prompts. Each entry averages 100 independent responses.

|  | Low Concern | Moderate Concern | High Concern |
|---|---|---|---|
| GMSTUX | 0.26 | 0.25 | 0.49 |
| AI Average | 0.00 | 0.00 | 1.00 |
| GPT-5 | 0.00 | 0.00 | 1.00 |
| GPT-4.1 | 0.00 | 0.00 | 1.00 |
| GPT-4o | 0.00 | 0.00 | 1.00 |
| GPT-4-Turbo | 0.00 | 0.00 | 1.00 |
| GPT-3.5-Turbo | 0.00 | 0.00 | 1.00 |
| Claude-Sonnet-4 | 0.00 | 0.00 | 1.00 |
| Claude-3.5-Haiku | 0.00 | 0.00 | 1.00 |
| Claude-3-Haiku | 0.00 | 0.00 | 1.00 |
| Gemini-2.5-Flash | 0.00 | 0.00 | 1.00 |
| Gemini-2.0-Flash | 0.00 | 0.00 | 1.00 |
| DeepSeek-V3.1 | 0.00 | 0.00 | 1.00 |
| Mistral-Large-2411 | 0.00 | 0.00 | 1.00 |

**Table 8 Summary Statistics for Firms with Analyst Recommendations**

This table reports presents the descriptive statistics of variables for firms with stock recommendations issued by financial analysts in our sample. The sample includes analysts from 19 brokerage firms, from which we collect reports published in 2023 and 2024. The recommendation sample contains analyst recommendations issued between 2020 and 2025. Variables are defined in Table A.1 in the Appendix.

|  | Obs | Mean | Std Dev | 5% | Median | 95% |
| --- | --- | --- | --- | --- | --- | --- |
| Recommendation Optimism | 17,350 | -0.254 | 0.811 | -2.000 | 0.000 | 1.000 |
| MSCI ESG Score | 17,350 | 4.451 | 1.366 | 2.031 | 4.5785 | 6.502 |
| LSEG ESG Score | 16,929 | 0.582 | 0.185 | 0.241 | 0.610 | 0.839 |
| After AI Usage | 17,350 | 0.024 | 0.155 | 0.000 | 0.000 | 0.000 |
| Total Assets | 17,350 | 22.40 | 1.610 | 19.77 | 22.41 | 25.09 |
| Leverage | 17,350 | 0.267 | 0.212 | 0.001 | 0.250 | 0.604 |
| ROA | 17,350 | 0.036 | 0.132 | -0.168 | 0.045 | 0.190 |
| Market-to-Book | 17,350 | 2.131 | 3.079 | 0.210 | 1.133 | 7.449 |
| Tangibility | 17,350 | 0.300 | 0.232 | 0.022 | 0.246 | 0.751 |
| Liquidity | 17,350 | 2.335 | 3.175 | 0.607 | 1.502 | 6.666 |
| Sales Growth | 17,350 | 0.171 | 0.624 | -0.234 | 0.078 | 0.695 |
| Market Share | 17,350 | 0.160 | 0.250 | 0.000 | 0.041 | 0.812 |

**Table 9 Corporate ESG Performance and Analyst Recommendations Around AI Usage**

This table reports the ESG performance of firms covered by sell-side analysts before and after their first use of LLM tools in research reports, separately by recommendation type. Recommendations are grouped into four categories: buy recommendations (Buy or Strong Buy), sell recommendations, recommendation upgrades, and recommendation downgrades. For each recommendation type, the table reports the number and the average ESG score of the firms. Panel A presents firms' MSCI ESG scores, and Panel B presents firms' LSEG ESG scores. The final two columns show the difference in mean ESG scores between the two regimes and the corresponding *t*-statistic from a two-sample mean comparison. ***, **, and * indicate significance at the 1%, 5%, and 10% levels, respectively.

Panel A: MSCI ESG Score

|  | After AI Usage | | Before AI Usage | | | |
| --- | --- | --- | --- | --- | --- | --- |
|  | Obs | Mean | Obs | Mean | Diff | *t*-stat |
| Buy | 258 | 4.519 | 580 | 4.166 | 0.352*** | 3.493 |
| Sell | 283 | 4.454 | 682 | 4.341 | 0.113 | 1.296 |
| Upgrade | 165 | 4.488 | 402 | 4.241 | 0.247** | 2.038 |
| Downgrade | 176 | 4.445 | 389 | 4.287 | 0.158 | 1.442 |

Panel B: LSEG ESG Score

|  | After AI Usage | | Before AI Usage | | | |
| --- | --- | --- | --- | --- | --- | --- |
|  | Obs | Mean | Obs | Mean | Diff | *t*-stat |
| Buy | 234 | 0.550 | 480 | 0.526 | 0.024* | 1.713 |
| Sell | 253 | 0.570 | 645 | 0.567 | 0.004 | 0.253 |
| Upgrade | 153 | 0.584 | 333 | 0.551 | 0.034** | 1.983 |
| Downgrade | 153 | 0.575 | 386 | 0.563 | 0.012 | 0.667 |

# Table 10 Analyst AI Usage and ESG-Related Recommendation Optimism

This table reports the impact of analysts' AI usage on their recommendations of stocks with different ESG performance. The dependent variable is *Recommendation Optimism*, defined as the difference between an analyst's stock recommendation (coded from 1 to 5, representing strong sell, sell, hold, buy, and strong buy, respectively) and the consensus recommendation—the median rating issued by all other analysts covering the same firm within the preceding 365 days. Higher values indicate more favorable recommendations. *Analyst AI Usage* is a dummy variable equal to one starting from the date when an analyst is first detected using AI assistance in research reports. *MSCI ESG Score* and *LSEG ESG Score* in Columns (1) and (2) are firms' ESG score obtained from MSCI and LESG, respectively. Controls include the natural logarithm of total assets (*Total Assets*), financial leverage ratio (*Leverage*), return on assets (*ROA*), the ratio of market value to book value (*Market-to-Book*), asset tangibility (*Tangibility*), the liquidity of firms' assets (*Liquidity*), annual growth rate of sales (*Sales Growth*), and firms' share of sales among all public firms in the same industry (*Market Share*). Variables are defined in Table A.1 in the Appendix. All specifications include firm, analyst, and time fixed effects. Standard errors are clustered at the analyst level, with *p*-values in parentheses. ***, **, and * indicate significance at the 1%, 5%, and 10% levels, respectively.

|  | (1) | (2) |
|---|---|---|
|  | \multicolumn{2}{c}{Recommendation Optimism} | |
| Analyst AI Usage × MSCI ESG Score | 0.0788** | |
|  | (0.03) | |
| Analyst AI Usage × LSEG ESG Score |  | 0.622** |
|  |  | (0.04) |
| Analyst AI Usage | -0.412** | -0.338 |
|  | (0.04) | (0.12) |
| MSCI ESG Score | -0.0122 |  |
|  | (0.16) |  |
| LSEG ESG Score |  | 0.0232 |
|  |  | (0.89) |
| Total Assets | -0.199*** | -0.191*** |
|  | (0.00) | (0.00) |
| Leverage | 0.0541 | 0.0464 |
|  | (0.65) | (0.70) |
| ROA | -0.0694 | 0.0484 |
|  | (0.55) | (0.66) |
| Market-to-Book | -0.0191*** | -0.0217*** |
|  | (0.00) | (0.00) |
| Tangibility | -0.127 | -0.0362 |
|  | (0.52) | (0.85) |
| Liquidity | 0.00783 | 0.00466 |
|  | (0.25) | (0.46) |
| Sales Growth | -0.0285* | -0.0269* |
|  | (0.07) | (0.06) |
| Market Share | -0.776*** | -0.791** |
|  | (0.01) | (0.01) |
| Constant | Yes | Yes |
| Firm Dummy | Yes | Yes |
| Analyst Dummy | Yes | Yes |
| Time Dummy | Yes | Yes |
| Cluster at Analyst Level | Yes | Yes |
| Observations | 17,350 | 16,929 |
| R-squared | 0.361 | 0.361 |

# Appendix

## Table A.1 Variable Definition

This table reports details on data sources and the definitions of variables used in our paper.

| Variable | Definition | Source |
| --- | --- | --- |
| Recommendation Optimism | The analyst's stock recommendation (coded as 1, 2, 3, 4, and 5 for strong sell, sell, hold, buy, and strong buy, respectively) minus the consensus recommendation — defined as the median recommendation of all other analysts covering the same firm within the 365-day window preceding the issuance of the focal recommendation (Ljungqvist et al., 2007). | I/B/E/S |
| MSCI ESG Score | MSCI ESG Score represents the weighted average of the Environmental, Social, and Governance pillar scores. | MSCI |
| LSEG ESG Score | LSEG ESG score measures the company's ESG performance based on verifiable reported data in the public domain. LSEG captures and calculates over 450 company-level ESG measures; a subset of 186 of the most comparable and material per industry (detailed in the ESG glossary, available on request) power the overall company assessment and scoring process. These are grouped into 10 categories that reformulate the three pillar scores and the final ESG score, which is a reflection of the company's ESG performance, commitment, and effectiveness based on publicly reported information. | Refinitiv ASSET4 |
| After AI Usage | A dummy variable that equals one if the recommendation is made after the release of an analyst report containing AI-related content, and zero otherwise. | Mergent, Pangram |
| Total Assets | Natural logarithm of raw total assets (Worldscope item 07230). Raw Total Assets represent the total assets of the company converted to U.S. dollars using the fiscal year-end exchange rate. | Worldscope |
| Leverage | Financial leverage ratio (Worldscope item 08236). Calculated as the ratio of total debt to total assets. Winsorized at level 1% and 99% levels. | Worldscope |
| ROA | Return on assets. Calculated as Net Income (Worldscope item 01651) / Total Assets (Worldscope item 02999). Winsorized at level 1% and 99% levels. | Worldscope |
| Market-to-Book | Market-to-book ratio. Calculated as Market Capitalization / (Total Assets - Total Liabilities), where Total Liabilities (Worldscope item 03351) represent all short- and long-term obligations expected to be satisfied by the company. A higher market-to-book tends to be a sign of more attractive future growth options, which a firm tends to protect by limiting its leverage. Winsorized at level 1% and 99% levels. | Worldscope |
| Tangibility | Tangibility of firms' assets. Calculated as Property, Plant and Equipment (Worldscope item 02501) / Total Assets (Worldscope item 02999). Property, Plant and Equipment represents Gross Property, Plant and Equipment less accumulated reserves for depreciation, depletion, and amortization. Firms operating with greater tangible assets have a higher debt capacity. Winsorized at level 1% and 99% levels. | Worldscope |
| Liquidity | Liquidity of firms' assets. Calculated as Total Current Assets (Worldscope item 02201) / Total Current Liabilities (Worldscope item 03101). Total Current Assets represents cash and other assets that are reasonably expected to be realized in cash, sold or consumed within one year or one operating cycle. Total Current Liabilities represent debt or other obligations that the company expects to satisfy within one year. Firms with more liquid assets can use them as another internal source of funds instead of debt, leading to lower optimal debt equity ratio. Winsorized at level 1% and 99% levels. | Worldscope |
| Sales Growth | The growth rate of firms' net sales, expressed in percentage (Worldscope item 08631). Calculated as Current Year's Net Sales or Revenues / Last Year's Total Net Sales or Revenues - 1. Winsorized at level 1% and 99% levels. | Worldscope |
| Market Share | Firm's share of sales by all public firms in the same Fama & French 48 industry and the same country. Winsorized at level 1% and 99% levels. | Worldscope |

# Internet Appendix

*Not for publication*

### Table IA.1 LLM Responses from Prompts without Role Definition

This table summarizes survey responses generated by GPT models when role-definition prompts are removed. All survey questions follow the formats used in Edmans et al. (2025) for professional active equity managers (EGJ) and Giglio et al. (GMSTUX, 2025) for U.S. retail investors. "GPT Average" reports the mean response across GPT model versions (i.e., GPT-5, GPT-4.1, GPT-4o, GPT-4-Turbo, and GPT-3.5-Turbo) using our baseline elicitation settings. "GPT Average (No Role)" reports the mean response across the same GPT models under prompts without role-definition instructions.

Panel A: Rank the following by their importance for the long-term value of companies in your investment universe in aggregate (1=most important, 6=least important).

|  | Strategy | Operational | Governance | Culture | Capital Structure | ES |
|---|---|---|---|---|---|---|
| EGJ | 1.67 | 2.36 | 3.71 | 4.12 | 4.13 | 5.01 |
| GPT Average | 1.18 | 2.53 | 2.70 | 4.11 | 5.95 | 4.53 |
| GPT Average (No Role) | 2.00 | 3.35 | 2.47 | 3.59 | 6.00 | 3.60 |

Panel B: How material is ES performance, on the following dimensions, to how you assess the long-term value of companies in your investment universe in aggregate? (0=immaterial, 4=highly material).

|  | Employee | Consumer | GHG | Pollution | Supplier | Ecology | Community | Diversity |  |
|---|---|---|---|---|---|---|---|---|---|
| EGJ | 2.59 | 2.53 | 2.50 | 2.49 | 2.31 | 2.23 | 1.99 | 1.68 | 2.29 |
| GPT Average | 3.14 | 3.05 | 3.22 | 3.06 | 2.25 | 3.01 | 2.28 | 1.99 | 2.75 |
| GPT Average (No Role) | 3.22 | 3.31 | 3.40 | 3.25 | 2.23 | 3.20 | 2.81 | 2.02 | 2.93 |

Panel C: Do you expect good ES performers (bad ES performers) to typically outperform or underperform in long-term risk-adjusted total shareholder return? (-2=strongly underperform, 0=neither under nor outperform, +2=strongly outperform).

|  | Good ES Performers | Bad ES Performers |
|---|---|---|
| EGJ | 0.57 | -0.70 |
| GPT Average | 1.27 | -1.27 |
| GPT Average (No Role) | 1.48 | -1.48 |

Panel D: How much long-term risk-adjusted total shareholder return would you tolerate a company sacrificing to improve its ES performance?

|  | Zero – I would not tolerate any sacrifice | 1-10 bp per year | 11-50 bp per year | >50 bp per year | No sacrifice is necessary since there is no trade-off |
|---|---|---|---|---|---|
| EGJ | 33% | 14% | 9% | 3% | 40% |
| GPT Average | 0% | 53% | 0% | 0% | 47% |
| GPT Average (No Role) | 0% | 49% | 0% | 0% | 51% |

Panel E: What do you expect the average annual return of the U.S. stock market (a diversified U.S. ESG equity portfolio) to be over the next 10 years? (Please answer only with a positive or negative numeric value with at most one decimal place.)

|  | Expected 10-year Annualized Return of US Stock Market | Expected 10-year Annualized Return of ESG Investments |
|---|---|---|
| GMSTUX | 7.13% | 5.20% |
| GPT Average | 6.71% | 6.61% |
| GPT Average (No Role) | 6.26% | 6.42% |

Panel F: Listed below are some reasons why individuals might invest in ESG portfolios. Please choose the one that you think is the most important for you.

|  | ESG will outperform | ESG hedges climate risk | It is the right thing to do | No specific reason |
|---|---|---|---|---|
| GMSTUX | 0.06 | 0.22 | 0.24 | 0.48 |
| GPT Average | 0.00 | 0.41 | 0.59 | 0.00 |
| GPT Average (No Role) | 0.00 | 0.44 | 0.56 | 0.00 |

Panel G: How do you describe your level of concern about climate change?

|  | Low Concern | Moderate Concern | High Concern |
|---|---|---|---|
| GMSTUX | 0.26 | 0.25 | 0.49 |
| GPT Average | 0.00 | 0.00 | 1.00 |
| GPT Average (No Role) | 0.00 | 0.00 | 1.00 |

**Table IA.2 ESG Content in Analyst Reports**

This table compares ESG-related content in analyst reports with and without AI-related components. Reports with AI content are defined as those containing measurable AI-generated elements, identified using the Pangram methodology. *Report with ESG Content* is a binary indicator equal to one if the report includes ESG-related language, and zero otherwise. *# ESG Sentences* denotes the number of ESG-related sentences within a given report. ESG-related content in analyst reports is identified by ChatGPT based on the following prompt: *Does the following text from an analyst report mention or discuss ESG or Sustainability topics? Respond with 'Yes' or 'No'. If the answer is 'Yes', also indicate how many sentences discuss ESG or Sustainability topics.* The table presents the mean values of these variables for both groups. The final two columns show the difference in means between reports with and without AI content, along with the corresponding *t*-statistics. ***, **, and * indicate significance at the 1%, 5%, and 10% levels, respectively.

|  | Reports with AI Content | | Reports without AI Content | | Diff | *t*-stat |
|---|---|---|---|---|---|---|
|  | Obs | Mean | Obs | Mean |  |  |
| Report with ESG Content | 105 | 0.695 | 32,411 | 0.355 | 0.340*** | 7.269 |
| # ESG Sentences | 105 | 4.371 | 32,411 | 1.333 | 3.038*** | 9.746 |